\documentclass[a4article,11pt]{article}
\usepackage{jheppub} 
\usepackage[T1]{fontenc}
\usepackage{booktabs}
\usepackage[dvips,table]{xcolor} 
\usepackage{xspace}
\usepackage{dcolumn}
\usepackage{hyperref}
\usepackage{caption}
\usepackage
[subrefformat=parens,position=top,skip=-15pt,margin=15pt,justification=justified,singlelinecheck=false]
{subcaption}

\allowdisplaybreaks[1]

\newcommand{\Pl}{\ell}
\newcommand{\fb}{{\ensuremath\unskip\,\text{fb}}\xspace}
\def\reffi#1{\mbox{Fig.~\ref{#1}}}
\def\reffis#1{\mbox{Figs.~\ref{#1}}}
\def\refta#1{\mbox{Table~\ref{#1}}}

\def\refse#1{\mbox{Sect.~\ref{#1}}}

\def\citere#1{\mbox{Ref.~\cite{#1}}}
\def\citeres#1{\mbox{Refs.~\cite{#1}}}

\def\be{\begin{equation}}
\def\ee{\end{equation}}

\newcommand{\PH}{\ensuremath{\text{H}}\xspace}
\newcommand{\Pj}{\ensuremath{\text{j}}\xspace}
\newcommand{\Pp}{\ensuremath{\text{p}}\xspace}
\newcommand{\Pe}{\ensuremath{\text{e}}\xspace}
\newcommand{\Pb}{\ensuremath{\text{b}}\xspace}
\newcommand{\Pq}{\ensuremath{\text{q}}\xspace}
\newcommand{\Pt}{\ensuremath{\text{t}}\xspace}
\newcommand{\Pu}{\ensuremath{\text{u}}\xspace}
\newcommand{\Pd}{\ensuremath{\text{d}}\xspace}
\newcommand{\Ps}{\ensuremath{\text{s}}\xspace}
\newcommand{\Pc}{\ensuremath{\text{c}}\xspace}
\newcommand{\Pg}{\ensuremath{\text{g}}\xspace}

\newcommand{\PW}{\ensuremath{\text{W}}\xspace}
\newcommand{\PZ}{\ensuremath{\text{Z}}\xspace}

\newcommand{\Mt}{\ensuremath{m_\Pt}\xspace}

\newcommand{\MWOS}{\ensuremath{M_\PW^\text{OS}}\xspace}
\newcommand{\MW}{\ensuremath{M_\PW}\xspace}
\newcommand{\MZOS}{\ensuremath{M_\PZ^\text{OS}}\xspace}
\newcommand{\MZ}{\ensuremath{M_\PZ}\xspace}

\newcommand{\Gt}{\ensuremath{\Gamma_\Pt}\xspace}
\newcommand{\GH}{\ensuremath{\Gamma_\PH}\xspace}

\newcommand{\GZOS}{\ensuremath{\Gamma_\PZ^\text{OS}}\xspace}

\newcommand{\GWOS}{\ensuremath{\Gamma_\PW^\text{OS}}\xspace}

\newcommand{\MVOS}{\ensuremath{M_V^\text{OS}}\xspace}%
\newcommand{\GVOS}{\ensuremath{\Gamma_V^\text{OS}}\xspace}%

\newcommand{\GeV}{\ensuremath{\,\text{GeV}}\xspace}
\newcommand{\TeV}{\ensuremath{\,\text{TeV}}\xspace}

\newcommand{\alphas}{\ensuremath{\alpha_\text{s}}\xspace}

\newcommand{\GF}{\ensuremath{G_\mu}}

\newcommand{\ptsub}[1]{\ensuremath{p_{\text{T},#1}}\xspace}

\newcommand{\etsub}[1]{\ensuremath{E_{\text{T},#1}}\xspace}

\newcommand{\recola}{{\sc Recola}\xspace}
\newcommand{\mocanlo}{{\sc MoCaNLO}\xspace}
\newcommand{\collier}{{\sc Collier}\xspace}
\newcommand{\madgraph}{{\sc\small MadGraph5\_aMC@NLO}\xspace}
\newcommand{\rT}{{\mathrm{T}}}
\newcolumntype{.}{D{.}{.}{-1}}
\newcolumntype{d}[1]{D{.}{.}{#1}}

\colorlet{tableoverheadcolor}{gray!37.5}
\colorlet{tableheadcolor}{gray!25}
\colorlet{tablerowcolor}{gray!12.5}

\newlength{\width}
\newlength{\height}

\def\draftdate{\relax}
\def\mda{\relax}
\def\mua{\relax}
\def\mla{\relax}
\def\draft{
\def\thtystars{******************************}
\def\sixtystars{\thtystars\thtystars}
\typeout{}
\typeout{\sixtystars**}
\typeout{* Draft mode!
         For final version remove \protect\draft\space in source file *}
\typeout{\sixtystars**}
\typeout{}
\def\draftdate{\today}
\def\mua{\marginpar[\boldmath\hfil$\uparrow$]%
                   {\boldmath$\uparrow$\hfil}\color{black}%
                    \typeout{marginpar: $\uparrow$}\ignorespaces}
\def\mda{\color{red}\marginpar[\boldmath\hfil$\downarrow$]%
                   {\boldmath$\downarrow$\hfil}%
                    \typeout{marginpar: $\downarrow$}\ignorespaces}
\def\mla{\marginpar[\boldmath\hfil$\rightarrow$]%
                   {\boldmath$\leftarrow $\hfil}%
                    \typeout{marginpar: $\leftrightarrow$}\ignorespaces}
\def\Mua{\marginpar[\boldmath\hfil$\Uparrow$]%
                   {\boldmath$\Uparrow$\hfil}\color{black}%
                    \typeout{marginpar: $\uparrow$}\ignorespaces}
\def\Mda{\color{red}\marginpar[\boldmath\hfil$\Downarrow$]%
                   {\boldmath$\Downarrow$\hfil}%
                    \typeout{marginpar: $\downarrow$}\ignorespaces}
\def\Mla{\marginpar[\boldmath\hfil\textcolor{red}{$\Rightarrow$}]%
                   {\boldmath\textcolor{red}{$\Leftarrow $}\hfil}%
                    \typeout{marginpar: $\leftrightarrow$}\ignorespaces}
\overfullrule 5pt
\oddsidemargin 15mm
\marginparwidth 29mm
}

\usepackage{color}

\title{Complete NLO corrections to $\PW^+\PW^+$ scattering and its irreducible background at the LHC}
\subheader{\today}

\author{Benedikt Biedermann,}
\author{Ansgar Denner,}
\author{Mathieu Pellen}

\affiliation{%
        Universit\"at W\"urzburg, %
        Institut f\"ur Theoretische Physik und Astrophysik, %
        Emil-Hilb-Weg 22, \linebreak %
        97074 W\"urzburg, %
        Germany%
}
\emailAdd{benedikt.biedermann@physik.uni-wuerzburg.de}
\emailAdd{ansgar.denner@physik.uni-wuerzburg.de}
\emailAdd{mathieu.pellen@physik.uni-wuerzburg.de}

\abstract{The process $\Pp\Pp\to\mu^+\nu_\mu\Pe^+\nu_{\Pe}\Pj\Pj$
  receives several contributions of different 
  orders in the strong and electroweak coupling constants. 
  Using appropriate event selections, this
  process is dominated by vector-boson scattering (VBS) and has
  recently been measured at the LHC.  It is thus of prime importance
  to estimate precisely each contribution.  In this article we compute
  for the first time the full NLO QCD and electroweak corrections to
  VBS and its irreducible background processes with realistic
  experimental cuts.  We do not rely on approximations but use
  complete amplitudes involving two different orders at tree
  level and three different orders at one-loop level.
  Since we take into account all interferences,
  at NLO level the corrections to the VBS process and to the
  QCD-induced irreducible background process contribute at the same
  orders.  Hence the two processes cannot be unambiguously
  distinguished, and all contributions to the
  $\mu^+\nu_\mu\Pe^+\nu_{\Pe}\Pj\Pj$ final state should be preferably
  measured together.}

\begin{document}

\maketitle
\flushbottom

\newpage

\section{Introduction}
\label{sec:introduction}

With the discovery of the Higgs boson in 2012 at the Large Hadron
Collider (LHC), a new era in particle physics has started. On the one
hand, this discovery substantiated the last missing ingredient of the
meanwhile well-established Standard Model of elementary particles. On
the other hand, it represents the dawn of a new paradigm of precision
physics aiming at the investigation of electroweak (EW) symmetry breaking.
Thereby, vector-boson scattering (VBS) plays a fundamental role owing
to its sensitivity to the quartic non-Abelian gauge couplings and to
the Higgs sector of the Standard Model as a whole.

At a hadron collider, the scattering of massive vector bosons occurs
if partons in the two incoming protons radiate W or Z bosons that
scatter off each other. The leptonic decay products of the scattered
bosons in association with two jets 
radiated from the incoming partons in forward direction 
give rise to a typical signature that can be
enhanced over the irreducible background with dedicated VBS event selections.
Among the various leptonic final states, the channel with two equally
charged leptons and two neutrinos, the so-called same-sign WW~channel,
has been identified as the most promising candidate for discovery
\cite{Campbell:2015vwa,Green:2016trm}. Owing to the limited number of
partonic channels that allow for such a leptonic final state, the
irreducible QCD background is smaller than in the other VBS channels.

During run~I at the LHC, evidence for VBS in the same-sign WW~channel
has been reported by both the ATLAS \cite{Aad:2014zda,Aaboud:2016ffv}
and CMS \cite{Khachatryan:2014sta} collaborations. Recently the CMS
collaboration has observed this process at the LHC with data from
run~II \cite{CMS:2017adb}.  It is therefore essential to have precise
and appropriate predictions for both the VBS process as well as for
its irreducible background.  In this context, \emph{precise} means
next-to-leading order (NLO) QCD and EW accuracy, and
\emph{appropriate} characterises predictions that are directly comparable with
experimental measurements.

In this article, we present the first complete computation of all NLO
QCD and EW corrections to the process
$\Pp\Pp\to\mu^+\nu_\mu\Pe^+\nu_{\Pe}\Pj\Pj$.
At leading order (LO), the cross section receives contributions of
three different orders in the EW and strong coupling constants
$\alpha$ and $\alphas$: 1)~a purely EW
contribution at $\mathcal{O}{\left(\alpha^{6}\right)}$ that includes
among others the actual VBS mechanism, 2)~a QCD-induced contribution
at $\mathcal{O}{\left(\alphas^{2}\alpha^{4}\right)}$, and 3)~an
interference contribution at $\mathcal{O}{\left(\alpha_{\rm
      s}\alpha^{5}\right)}$. While the purely EW contribution contains,
besides the VBS contribution, also irreducible background
contributions and triple W-boson production, we will nevertheless sometimes
refer to the EW production mode as VBS process in the following.  From
a theoretical point of view, the three LO contributions can be
separated in a gauge-invariant way based on the different orders in
the coupling constants. 
From an experimental point of view, dedicated
phase-space cuts, including tagging jets with large rapidity separation
and invariant mass, have been designed to enhance
the actual VBS contribution from its QCD-induced
and EW-induced irreducible background.

Consequently, the complete NLO contribution involves the four
different orders $\mathcal{O}{\left(\alpha^{7}\right)}$,
$\mathcal{O}{\left(\alphas\alpha^{6}\right)}$,
$\mathcal{O}{\left(\alphas^{2}\alpha^{5}\right)}$, and
$\mathcal{O}{\left(\alphas^{3}\alpha^{4}\right)}$. Since some
of these single NLO contributions furnish corrections to more than one
LO contribution, it is not possible to unambiguously
attribute a given type of correction to a given underlying Born
process. Hence, at NLO one cannot distinguish the different production
mechanisms, in particular EW- and QCD-induced production modes, as they
are naturally mutually contaminated.

Parts of the NLO corrections to the process
$\Pp\Pp\to\mu^+\nu_\mu\Pe^+\nu_{\Pe}\Pj\Pj$ have already been
computed 
in the literature.  
These calculations focused on NLO QCD corrections for both
the VBS process~\cite{Jager:2009xx,Jager:2011ms,Denner:2012dz,
  Rauch:2016pai} and its QCD-induced irreducible background
process~\cite{Melia:2010bm,Melia:2011gk,Campanario:2013gea,Baglio:2014uba,Rauch:2016pai}.
We have already reported in \citere{Biedermann:2016yds} on the
surprisingly large NLO EW corrections to the VBS process.
The aim of the present article is to provide the complete NLO
corrections to the $\mu^+\nu_\mu\Pe^+\nu_{\Pe}\Pj\Pj$ final state,
based on the complete LO and NLO matrix elements and including all
interference contributions and all off-shell effects.

This article is organised as follows. In \refse{sec:calculation},
details of the calculation are described. In particular, the different
types of real and virtual corrections, and the 
validation of our results
are reviewed.  In \refse{sec:results}, 
numerical results are presented for integrated cross sections and
differential distributions.  The article concludes with a
summary and final remarks in \refse{sec:Conclusions}.

\section{Details of the calculation}
\label{sec:calculation}

\begin{figure}[t]
\begin{center}
          \includegraphics[width=0.30\linewidth]{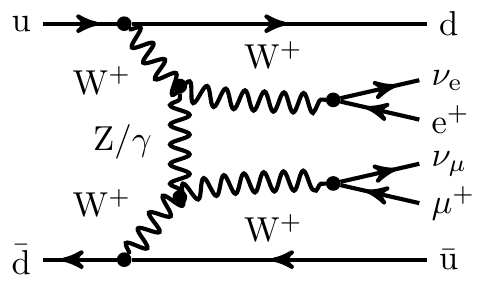}
          \includegraphics[width=0.30\linewidth]{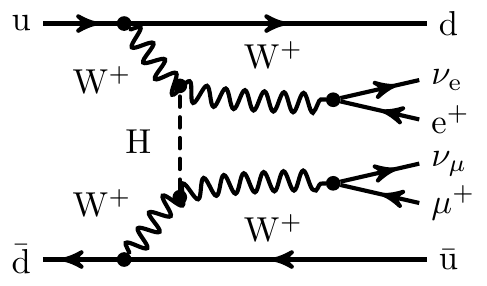}
          \includegraphics[width=0.30\linewidth]{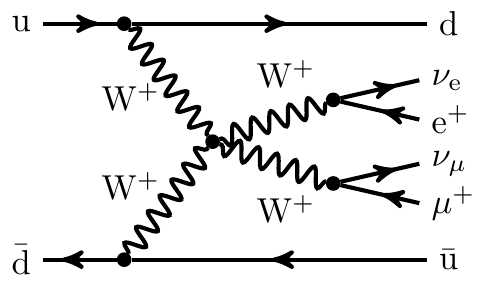}\\[2ex]
          \raisebox{.5ex}{\includegraphics[width=0.35\linewidth]{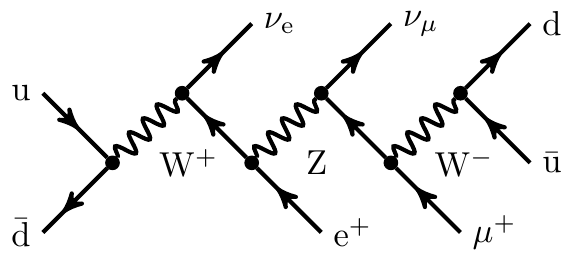}}
          \raisebox{-1.8ex}{\includegraphics[width=0.32\linewidth]{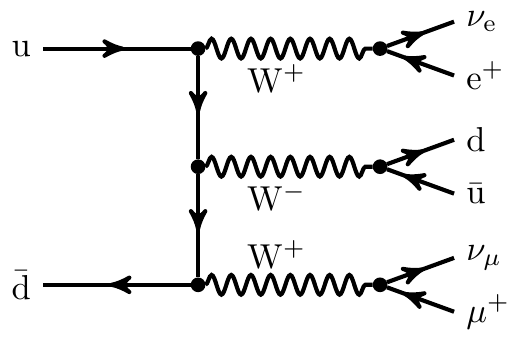}}
          \raisebox{.1ex}{\includegraphics[width=0.30\linewidth]{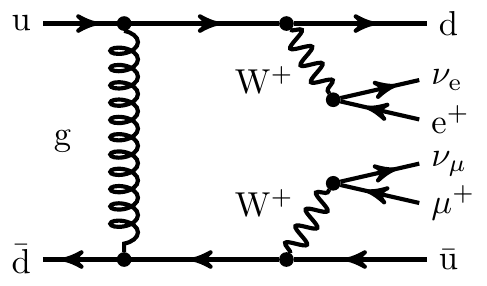}}
\end{center}
        \caption{Sample tree-level diagrams that contribute to the
          process
          $\Pp\Pp\to\mu^+\nu_\mu\Pe^+\nu_{\Pe}\Pj\Pj$.}
\label{diag:LO}
\end{figure}

The hadronic process studied is defined  at LO as
\begin{equation}
 \Pp\Pp\to\mu^+\nu_\mu\Pe^+\nu_{\Pe}\Pj\Pj .\label{eq:LOhadproc}
\end{equation}
Owing to charge conservation, there are no gluon-induced or
photon-induced contributions at LO. Furthermore, bottom quarks in the
initial state do not contribute as these would lead to a final state
with massive top quarks which falls under a different experimental
signature.  At the amplitude level, the process receives two different
types of contributions: a pure EW part at the order
$\mathcal{O}{\left(g^{6}\right)}$ (which we call sometimes simply VBS
contribution) and a QCD-induced part at the order
$\mathcal{O}{\left(g_{\rm s}^2 g^{4}\right)}$ with $g$ and $g_{\rm s}$
being the EW and QCD coupling constants, respectively.
Figure~\ref{diag:LO} shows sample tree-level diagrams for the partonic
sub-process $\Pu\bar\Pd\to\mu^+\nu_\mu\Pe^+\nu_{\Pe}\bar\Pu\Pd$.  The
top row of diagrams illustrates the actual VBS process at
$\mathcal{O}{\left(g^{6}\right)}$ with its characteristic VBS topology
of two $\PW$ bosons 
with space-like momenta
 that scatter into two
$\PW$~bosons 
with time-like momenta.
These contributions are referred to as $t$-channel diagrams since the
two incoming quark/anti-quark lines are connected to outgoing
quark/anti-quark lines.  For identical outgoing quarks or anti-quarks
also $u$-channel diagrams are obtained by exchanging the two outgoing
quarks or anti-quarks.  The $s$-channel diagram on the left in the
bottom row of order $\mathcal{O}{\left(g^{6}\right)}$ contributes to
the irreducible EW background.  In general, $s$-channel diagrams are
diagrams where the incoming quark and anti-quark are connected via
fermion lines.  There are also $s$-channel diagrams contributing to
triple gauge-boson production ($\PW^+\PW^+\PW^-$) (bottom middle).
Finally, the
diagram on the bottom right is an example of a QCD-induced contribution at
order $\mathcal{O}{\left(g_{\rm s}^2 g^{4}\right)}$.  This
contribution exclusively consists of diagrams where a gluon is
connecting the two quark lines and thus, by construction, cannot
involve VBS topologies.  Thus, at the level of squared amplitudes,
three gauge-invariant contributions exist: the pure EW contribution of
order $\mathcal{O}{\left(\alpha^{6}\right)}$, the QCD-induced
contribution of order $\mathcal{O}{\left(\alpha_{\rm
      s}^{2}\alpha^{4}\right)}$, and 
interferences of the order
$\mathcal{O}{\left(\alphas\alpha^{5}\right)}$.  
Owing to the colour structure, these interferences
occur only if diagrams of different quark flow between initial and
final state are multiplied with each other.
Thus, order-$\mathcal{O}{\left(\alphas\alpha^{5}\right)}$ contributions appear
only in partonic channels that involve contributions of two different
kinematic channels ($s$, $t$, $u$).
For example, in \reffi{diag:LO}, the contraction of the QCD-induced
diagram (bottom right) with the 
VBS diagrams (top row) necessarily vanishes due
to colour structure, while the corresponding contraction with the EW
$s$-channel background diagrams (bottom left and bottom middle) 
leads to a non-zero interference contribution at order $\mathcal{O}{\left(\alpha_{\rm s}\alpha^{5}\right)}$.
We stress that we include in our
calculation all possible contributions at the orders
$\mathcal{O}{\left(\alpha^{6}\right)}$, $\mathcal{O}{\left(\alpha_{\rm
        s}\alpha^{5}\right)}$, and $\mathcal{O}{\left(\alpha_{\rm
      s}^{2}\alpha^{4}\right)}$ that belong to the hadronic process in
Eq.~(\ref{eq:LOhadproc}).  A list of all contributing independent partonic
channels is given in
\refta{tab:partonic-channels},
which provides also information on contributing kinematic channels and
interferences.
\begin{table}[b]
\begin{center}
\begin{tabular}{|l@{}l|c|c|}
\hline
\multicolumn{2}{|c|}{partonic channel} & interferences at
$\mathcal{O}{\left(\alphas\alpha^{5}\right)}$ & kinematic channels\\
\hline
$ \Pu   \Pu        $&${}\to \mu^+ \nu_\mu  \Pe^+ \nu_\Pe  \Pd   \Pd$ &
yes & $t,u$\\
$ \Pu   \Pc/\Pc   \Pu      $&${}\to \mu^+ \nu_\mu  \Pe^+ \nu_\Pe  \Pd   \Ps$ & no  & $t$\\
$ \Pc   \Pc        $&${}\to \mu^+ \nu_\mu  \Pe^+ \nu_\Pe  \Ps   \Ps$ & yes  & $t,u$\\
$ \Pu   \bar\Pd/\bar\Pd  \Pu  $&${}\to \mu^+ \nu_\mu  \Pe^+ \nu_\Pe \Pd \bar\Pu$ & yes & $t,s$ \\
$ \Pu   \bar\Pd/\bar\Pd\Pu    $&${}\to \mu^+ \nu_\mu  \Pe^+ \nu_\Pe \Ps \bar\Pc $ & no  & $s$\\
$ \Pu   \bar\Ps/\bar\Ps  \Pu  $&${}\to \mu^+ \nu_\mu  \Pe^+ \nu_\Pe \Pd \bar\Pc  $ & no & $t$ \\
$ \Pc   \bar\Pd/\bar\Pd  \Pc  $&${}\to \mu^+ \nu_\mu  \Pe^+ \nu_\Pe \Ps \bar\Pu  $ & no  & $t$\\
$ \Pc \bar\Ps/ \bar\Ps  \Pc  $&${}\to \mu^+ \nu_\mu  \Pe^+ \nu_\Pe \Pd\bar\Pu $ & no  & $s$\\
$ \Pc \bar\Ps/ \bar\Ps  \Pc  $&${}\to \mu^+ \nu_\mu  \Pe^+ \nu_\Pe \Ps\bar\Pc  $ & yes  & $t,s$\\
$ \bar\Pd  \bar\Pd $&${}\to \mu^+ \nu_\mu  \Pe^+ \nu_\Pe  \bar\Pu  \bar\Pu$ & yes  & $t,u$\\
$ \bar\Pd  \bar\Ps/\bar\Ps  \bar\Pd $&${}\to \mu^+ \nu_\mu  \Pe^+ \nu_\Pe  \bar\Pu  \bar\Pc$ & no & $t$ \\
$ \bar\Ps  \bar\Ps $&${}\to \mu^+ \nu_\mu  \Pe^+ \nu_\Pe  \bar\Pc  \bar\Pc$ & yes  & $t,u$\\
\hline
\end{tabular}
\end{center}
\caption{Leading-order partonic channels contributing to the hadronic
  process $\Pp\Pp\to\mu^+\nu_\mu\Pe^+\nu_{\Pe}\Pj\Pj$. The middle column
   indicates whether the channel gives rise to an
  interference contribution at $\mathcal{O}{\left(\alpha_{\rm
        s}\alpha^{5}\right)}$ or not. The right column specifies
  the contributing kinematic channels.}\label{tab:partonic-channels} 
\end{table}

\begin{figure}[t]
\begin{center}
          \includegraphics[width=1.0\linewidth]{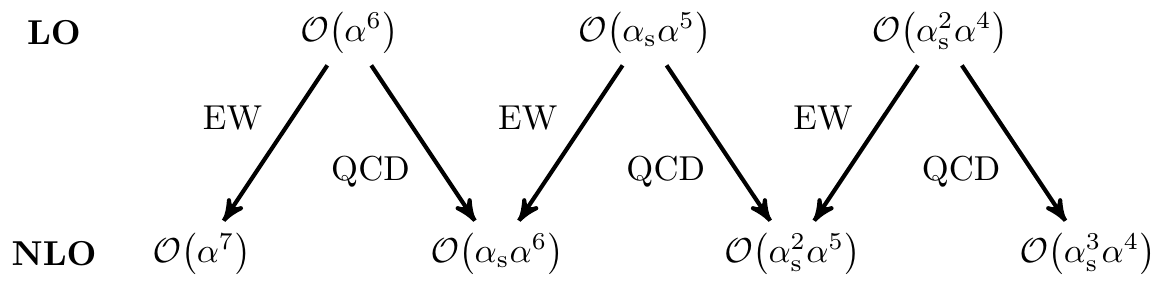}
\end{center}
        \caption{All contributing orders at both LO and NLO for the process $\Pp\Pp\to\mu^+\nu_\mu\Pe^+\nu_{\Pe}\Pj\Pj$.}
\label{allorders}
\end{figure}

At NLO, we compute both the QCD and EW corrections to each LO
contribution.  This leads to four possible NLO orders:
$\mathcal{O}{\left(\alpha^{7}\right)}$,
$\mathcal{O}{\left(\alphas\alpha^{6}\right)}$,
$\mathcal{O}{\left(\alphas^{2}\alpha^{5}\right)}$, and
$\mathcal{O}{\left(\alphas^{3}\alpha^{4}\right)}$.  The situation is
represented graphically in \reffi{allorders}.%
\footnote{Such a classification in
powers of $\alphas$ and $\alpha$ can also be found in
\citere{Alwall:2014hca}.}
The order $\mathcal{O}{\left(\alpha^{7}\right)}$ contributions are
simply the NLO EW corrections to the EW-induced LO processes. They
have already been presented in \citere{Biedermann:2016yds} for a fixed
scale.  Similarly, the order $\mathcal{O}{\left(\alpha_{\rm
      s}^{3}\alpha^{4}\right)}$ contributions furnish the QCD
corrections to the QCD-induced process, which have been computed in
\citeres{Melia:2010bm,Melia:2011dw,Campanario:2013gea}.

For the orders $\mathcal{O}{\left(\alphas\alpha^{6}\right)}$
and $\mathcal{O}{\left(\alphas^{2}\alpha^{5}\right)}$, a simple
separation of the EW-induced process and the QCD-induced process is not
possible any more, also for the dominant $\Pu\Pu$ 
partonic
channel.  Indeed,
the order $\mathcal{O}{\left(\alphas\alpha^{6}\right)}$
contains QCD corrections to the VBS process as well as EW corrections
to the LO interference.
The QCD corrections have already been
computed in the VBS approximation in
\citeres{Jager:2009xx,Jager:2011ms,Denner:2012dz,Campanario:2013gea,Baglio:2014uba}.
This means that the $s$-channel diagrams as well as the interference
of $t$- and $u$-channel diagrams are neglected.  In this
approximation, the interferences of the LO VBS and QCD-induced
contribution are vanishing.  Similarly, the order
$\mathcal{O}{\left(\alphas^{2}\alpha^{5}\right)}$ contains EW
corrections to the QCD-induced contribution as well as QCD corrections
to the LO interference.  These corrections have never been computed
previously and are presented here for the first time.

All the tree-level and one-loop matrix elements have been obtained
from the computer code \recola~\cite{Actis:2012qn,Actis:2016mpe} based
on the \collier~\cite{Denner:2014gla,Denner:2016kdg} library.
Throughout, the complex-mass scheme \cite{Denner:1999gp,Denner:2005fg}
is used.  All results have been obtained in two independent Monte
Carlo programs that have already been used for the computations of NLO
QCD and EW corrections for high-multiplicity processes described in
\citeres{Biedermann:2016yvs,Biedermann:2016guo,Biedermann:2016yds,Biedermann:2016lvg}
and
\citeres{Denner:2015yca,Denner:2016jyo,Biedermann:2016yds,Denner:2016wet},
respectively.

\subsection{Real corrections}
\label{sec:RealCorrections}

In this section, the real NLO corrections (both of QCD and QED origin) 
to all LO contributions are discussed.  
To handle the associated IR divergences,
the dipole-subtraction method for QCD \cite{Catani:1996vz} and its
extension to QED \cite{Dittmaier:1999mb} have been employed.  The
colour-correlated matrix elements needed for the subtraction procedure
are obtained directly from \recola.

At the order $\mathcal{O}{\left(\alpha^{7}\right)}$, the real
corrections consist simply of all photon radiations off any charged
particle, \emph{i.e.}\ all contributions of the type
$\Pp\Pp\to\mu^+\nu_\mu\Pe^+\nu_{\Pe}\Pj\Pj\gamma$ originating from the
LO EW production mode.
At the order $\mathcal{O}{\left(\alphas\alpha^{6}\right)}$,
there are two types of real radiation.  First, there is the QCD
radiation $\Pp\Pp\to\mu^+\nu_\mu\Pe^+\nu_{\Pe}\Pj\Pj\Pj$ with
underlying EW Born.  Second, there is photon radiation
$\Pp\Pp\to\mu^+\nu_\mu\Pe^+\nu_{\Pe}\Pj\Pj\gamma$ from the LO
interferences.  While both types of real radiation contribute at order
$\mathcal{O}{\left(\alphas\alpha^{6}\right)}$, each type
requires a different subtraction procedure.
In the same way, the order $\mathcal{O}{\left(\alphas^{2}\alpha^{5}\right)}$ features two types of real contributions.
First, photon radiation from the QCD-induced process and, second, QCD
radiation from the LO interference contributions.
Finally, the QCD radiation to the QCD-induced process of the type
$\Pp\Pp\to\mu^+\nu_\mu\Pe^+\nu_{\Pe}\Pj\Pj\Pj$ contributes at the
order $\mathcal{O}{\left(\alphas^{3}\alpha^{4}\right)}$.

Note that the QCD radiation of type
$\Pp\Pp\to\mu^+\nu_\mu\Pe^+\nu_{\Pe}\Pj\Pj\Pj$ includes both gluon
radiation from any coloured particle as well as quark/anti-quark
radiation from $\Pg\bar{\Pq}$ and $\Pg\Pq$ initial states. The
corresponding partonic channels can systematically be obtained from
the list of partonic channels at LO given in
\refta{tab:partonic-channels} by first attaching an additional gluon
to the final state, and then crossing this gluon with one of the
quarks or anti-quarks in the initial state.  In the same way, real
radiation from photon-induced contributions of the type $\gamma \Pq /
\gamma \bar{\Pq} \to\mu^+\nu_\mu\Pe^+\nu_{\Pe}\Pj\Pj\Pj$ contributes at
the orders $\mathcal{O}{\left(\alpha^{7}\right)}$,
$\mathcal{O}{\left(\alphas\alpha^{6}\right)}$, and
$\mathcal{O}{\left(\alphas^{2}\alpha^{5}\right)}$.  We have
computed these contributions separately (\emph{c.f.}\ 
\refta{table:photoninducedXsection}) but do not include them in our
default NLO corrections.

\subsection{Virtual corrections}
\label{ssec:VirtualCorrections}

\begin{figure}
\begin{center}
          \includegraphics[width=0.3\linewidth]{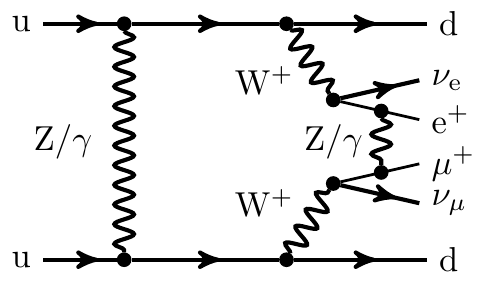}
          \includegraphics[width=0.34\linewidth]{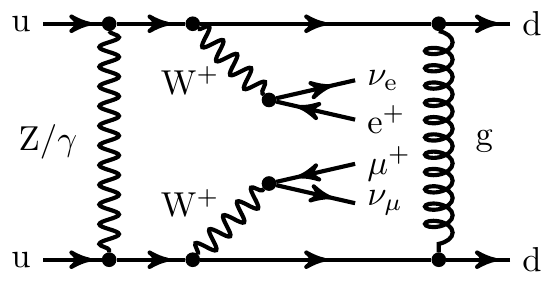}
          \includegraphics[width=0.3\linewidth]{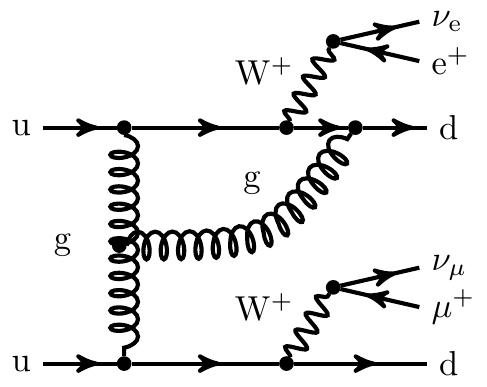}
\end{center}
        \caption{Sample one-loop level diagrams contributing to the process $\Pp\Pp\to\mu^+\nu_\mu\Pe^+\nu_{\Pe}\Pj\Pj$.}
\label{diag:NLO}
\end{figure}

In the same way as for the real corrections, the various virtual
corrections contributing at each order are described in the
following.  All the virtual corrections have been obtained from
\recola in association with the \collier library which is used to
calculate the one-loop scalar
\cite{tHooft:1978jhc,Beenakker:1988jr,Dittmaier:2003bc,Denner:2010tr}
and tensor integrals
\cite{Passarino:1978jh,Denner:2002ii,Denner:2005nn} numerically.  Some
of the virtual diagrams computed are represented in \reffi{diag:NLO}.
On the left-hand side, an EW correction to the EW amplitude, a
diagram of order $\mathcal{O}{\left(g^{8}\right)}$ featuring an
8-point function, is shown.
The diagram of order $\mathcal{O}{\left(g_{\rm s}^{2}g^{6}\right)}$ in the middle can either be interpreted as
an EW correction to the QCD-induced process or as a QCD correction to
the EW-induced process and illustrates that both processes cannot be 
separated any more once the full NLO corrections are included.
Finally, on the right hand-side a QCD loop correction to the QCD-induced
amplitude is displayed.

In order to understand the emergence of different orders in the EW and
QCD coupling at the level of the virtual corrections, one starts again
at amplitude level and considers all possible interferences of the
Born contributions at order $\mathcal{O}{\left(g^{6}\right)}$ and
$\mathcal{O}{\left(g_{\rm s}^2 g^{4}\right)}$ with the virtual
amplitudes at the orders $\mathcal{O}{\left(g^{8}\right)}$,
$\mathcal{O}{\left(g_{\rm s}^2 g^{6}\right)}$, and
$\mathcal{O}{\left(g_{\rm s}^4 g^{4}\right)}$.
At the order $\mathcal{O}{\left(\alpha^{7}\right)}$, the virtual
corrections consist simply of EW corrections to the EW tree-level
amplitude interfered with the EW tree-level amplitude.
\begin{figure}[t]
\begin{center}
          \includegraphics[width=0.9\linewidth]{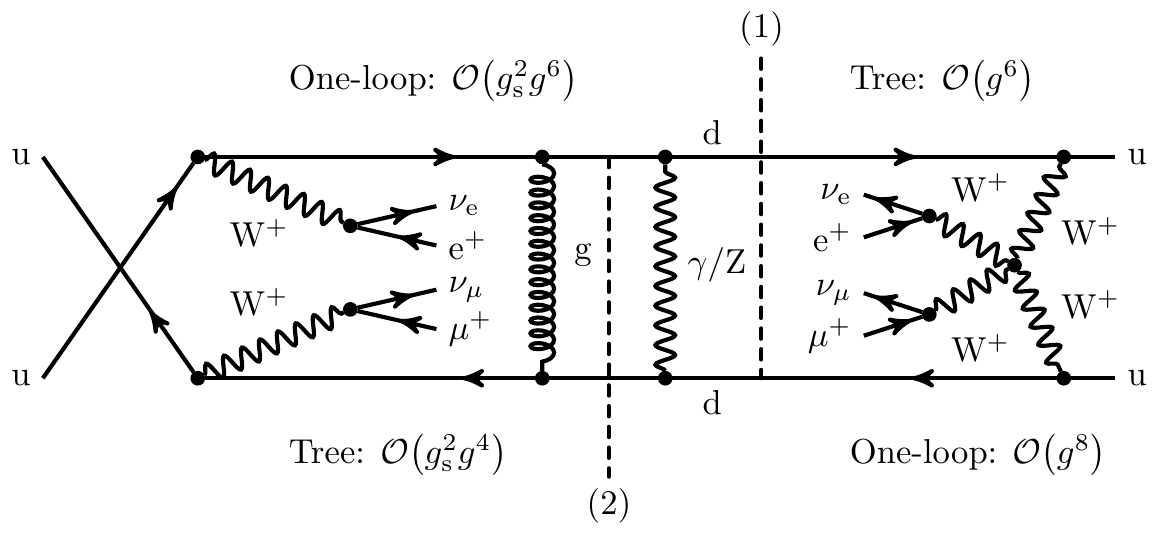}
\end{center}
        \caption{Contribution to the squared matrix element at the
          order $\mathcal{O}{\left(\alphas\alpha^{6}\right)}$. 
        It can be viewed as an amplitude of order
        $\mathcal{O} \left( g^2_{\rm s} g^6 \right)$ interfered with
        the LO EW amplitude [cut (1)].
        On the other hand, it can be seen as an EW correction to the EW amplitude
        interfered with the LO QCD amplitude [cut (2)].
        Owing to the colour structure, the illustrated contractions
        necessarily connect $t$- and $u$-channel contributions.}         
\label{fig:SquaredMatrix}
\end{figure}
Concerning the order $\mathcal{O}{\left(\alpha_{\rm s}\alpha^{6}\right)}$, there are different types of
contributions. One first considers the insertions of gluons into the
purely EW tree-level amplitude as well as the EW corrections to the QCD-induced tree-level
amplitude leading to a one-loop amplitude at $\mathcal{O}{\left(g_{\rm s}^2 g^{6}\right)}$ (see middle diagram of \reffi{diag:NLO} for a diagrammatic example).
This one-loop amplitude is then interfered with the EW tree-level
amplitude at $\mathcal{O}{\left(g^{6}\right)}$.
The contraction is illustrated at the level of squared amplitudes in Fig.~\ref{fig:SquaredMatrix} via the cut along the dashed line number (1).
Second, the EW corrections to the EW tree-level amplitude at $\mathcal{O}{\left(g^{8}\right)}$
contracted with the QCD-induced LO amplitude at $\mathcal{O}{\left(g_{\rm s}^2 g^{4}\right)}$ lead to yet another
contribution of order $\mathcal{O}{\left(\alpha_{\rm s}\alpha^{6}\right)}$. This corresponds to the cut along the dashed line number (2) in Fig.~\ref{fig:SquaredMatrix}.
While real photon radiation and real QCD radiation are still distinguishable at the level of squared amplitudes, Fig.~\ref{fig:SquaredMatrix} illustrates that
this is not the case any more for the virtual corrections.
The situation is similar at the order $\mathcal{O}{\left(\alpha_{\rm s}^{2}\alpha^{5}\right)}$.
First, there are interferences of the
QCD-induced tree-level amplitude with QCD corrections to the
EW-induced LO amplitude and the EW corrections to the QCD-induced LO
amplitude.  Second, the gluon insertions in the QCD-induced tree-level
amplitude, interfered with the EW-induced tree-level amplitude lead to
another contribution at order $\mathcal{O}{\left(\alpha_{\rm s}^{2}\alpha^{5}\right)}$.
Finally, the virtual contributions of order
$\mathcal{O}{\left(\alphas^{3}\alpha^{4}\right)}$ consist
simply of the QCD corrections to the QCD-induced tree-level amplitude
contracted with the QCD-induced tree-level amplitude.

\subsection{Validation}
\label{sec:Validations}

The computation has been done with two different Monte Carlo programs
providing thus an independent check of
the phase-space integration and the treatment of IR singularities.
These two Monte Carlo programs have already shown to be reliable when
computing both NLO QCD and EW corrections for a variety of processes
\cite{Denner:2015yca,Biedermann:2016yvs,Biedermann:2016guo,Biedermann:2016yds,Biedermann:2016lvg,Denner:2016jyo,Denner:2016wet}.
The photon-induced contributions have been implemented in one of these
Monte Carlo codes, but have been cross-checked with independent
programs for vector-boson pair production
\cite{Biedermann:2016guo,Biedermann:2016lvg}.  While all amplitudes
are obtained from \recola, the computation has been performed with two
different libraries of the
\collier~\cite{Denner:2014gla,Denner:2016kdg} program, apart from the
purely EW virtual amplitudes.  The results obtained at the integration
level are in excellent numerical agreement demonstrating thus the
stability of the virtual contribution.  The virtual corrections of
order $\mathcal{O}{\left(\alphas^{3}\alpha^{4}\right)}$ for the
$\Pu\Pu$~channel obtained from \recola agree within at least 6 digits
with the ones obtained with {\sc MadLoop} \cite{Hirschi:2011pa}, part
of the \madgraph \cite{Alwall:2014hca} framework for more than $99\%$
of 3000 phase-space points generated randomly.  Finally, we recall
that the NLO computation at the order
$\mathcal{O}{\left(\alpha^{7}\right)}$ reported in
\citere{Biedermann:2016yds} (computed with fixed scale) has already
undergone several validations.
These comprise a phase-space point comparison of representative
tree-level matrix elements squared and a comparison for the dominant
channels against an NLO double-pole approximation.  The implementation
of the double-pole approximation has been automatised and used in
\citeres{Denner:2016jyo,Denner:2016wet}.

\section{Numerical Results}
\label{sec:results}

\subsection{Input parameters and selection cuts}
\label{ssec:InputParameters}

The results presented are for the LHC operating at a centre-of-mass
energy of $13\TeV$.  
As parton distribution functions (PDF) we use the
NLO NNPDF-3.0 QED set with $\alphas(\MZ) =
0.118$ \cite{Ball:2013hta,Ball:2014uwa}, 
interfaced to our Monte Carlo
programs via LHAPDF~6.1.5~\cite{Andersen:2014efa,Buckley:2014ana}.  
We have employed the fixed $N_\text{F}=5$ flavour scheme throughout.
The EW collinear initial-state splittings are handled within the DIS
factorisation scheme \cite{Diener:2005me,Dittmaier:2009cr}, while the
QCD ones are treated by ${\overline{\rm MS}}$ redefinition of the PDF.
We use the same PDF for LO and NLO predictions.
The renormalisation and factorisation scales are set dynamically as
\begin{equation}
\label{eq:defscale}
 \mu_{\rm ren} = \mu_{\rm fac} = \sqrt{p_{\rm T, j_1}\, p_{\rm T, j_2}},
\end{equation}
where $p_{\rm T, j_i}$, $i=1,2$, are the transverse momenta of the two leading
jets (see below for the definition).  This scale has been found to
reduce significantly the difference between LO and NLO QCD predictions
for the VBS process at large transverse momenta \cite{Denner:2012dz}.

Regarding the electromagnetic coupling, the $G_\mu$ scheme \cite{Denner:2000bj} has been used where the coupling is obtained from the Fermi constant as
\begin{equation}
  \alpha = \frac{\sqrt{2}}{\pi} G_\mu \MW^2 \left( 1 - \frac{\MW^2}{\MZ^2} \right)  \qquad \text{with}  \qquad   {\GF    = 1.16637\times 10^{-5}\GeV^{-2}}.
\end{equation}
The masses and widths of the massive particles read \cite{Agashe:2014kda}
\begin{alignat}{2}
\label{eqn:ParticleMassesAndWidths}
                  \Mt   &=  173.21\GeV,       & \quad \quad \quad \Gt &= 0 \GeV,  \nonumber \\
                \MZOS &=  91.1876\GeV,      & \quad \quad \quad \GZOS &= 2.4952\GeV,  \nonumber \\
                \MWOS &=  80.385\GeV,       & \GWOS &= 2.085\GeV,  \nonumber \\
                M_{\rm H} &=  125.0\GeV,       &  \GH   &=  4.07 \times 10^{-3}\GeV.
\end{alignat}
The bottom quark is considered massless and does not appear in the initial state for the process under consideration.
The width of the top quark is set to zero as it is never resonant.
The Higgs-boson mass is taken according to the recommendation of the Higgs cross section working group \cite{Heinemeyer:2013tqa} with its corresponding width.
The pole masses and widths entering the calculation are determined
from the measured on-shell (OS) values \cite{Bardin:1988xt} for the W and Z~bosons according to
\begin{equation}
        M_V = \frac{\MVOS}{\sqrt{1+(\GVOS/\MVOS)^2}}\,,\qquad  
\Gamma_V = \frac{\GVOS}{\sqrt{1+(\GVOS/\MVOS)^2}}.
\end{equation}

The set of acceptance cuts that we employ is inspired from
\citeres{Aad:2014zda,Khachatryan:2014sta,CMS:2017adb} which describe
searches for the VBS process at the LHC at a centre-of-mass energy of
$8\TeV$ and $13\TeV$.
Experimentally, the final state of the process is required to have two
equally charged leptons, missing transverse energy and at least two
jets. QCD partons are clustered into jets using the anti-$k_\text{T}$
algorithm \cite{Cacciari:2008gp} with jet-resolution parameter
$R=0.4$.  Similarly, photons from real radiation are recombined with
the final-state quarks into jets or with the charged leptons into
dressed leptons, in both cases via the anti-$k_\text{T}$ algorithm and
a resolution parameter $R=0.1$.  Only partons with rapidity $|y|<5$
are considered for recombination, while particles with larger $|y|$ are
assumed to be lost in the beam pipe.  The rapidity $y$ and the
transverse momentum $p_{\rm T}$ of a particle are defined as
\begin{align}
y=\frac{1}{2}\ln \frac{E+p_z}{E-p_z}, \qquad
p_{\rm T} = \sqrt{p_x^2+p_y^2},
\end{align}
where $E$ is the energy of the particle, $p_z$ the component of its
momentum along the beam axis, and $p_x$, $p_y$ the components
perpendicular to the beam axis.  

The charged leptons $\ell$ are required to fulfil the acceptance
cuts
\begin{align}
 \ptsub{\Pl} >  20\GeV,\qquad |y_{\Pl}| < 2.5, \qquad \Delta R_{\Pl\Pl}> 0.3.
\end{align}
The distance $\Delta R_{ij}$ between two particles $i$ and $j$ in the rapidity--azimuthal-angle plane reads
\begin{equation}
        \Delta R_{ij} = \sqrt{(\Delta \phi_{ij})^2+(\Delta y_{ij})^2},
\end{equation}
with $\Delta \phi_{ij}=\min(|\phi_i-\phi_j|,2\pi-|\phi_i-\phi_j|)$
being the azimuthal-angle difference and $\Delta y_{ij} = y_i-y_j$ the
rapidity difference. The missing transverse energy is required to
fulfil
\begin{align}
  \etsub{\text{miss}}=p_{\rm T, miss} >  40\GeV
\end{align}
and is computed as the transverse momentum of the sum of the two
neutrino momenta. A QCD parton system after recombination is called a jet if
it obeys the jet-identification criteria
\begin{align}
 \ptsub{\Pj} >  30\GeV, \qquad |y_\Pj| < 4.5,\qquad\Delta R_{\Pj\Pl} > 0.3,
\end{align}
where the last condition requires a minimal distance between a jet and
each of the charged leptons. The identified jets are then ordered according to the size of their transverse momenta.
On the invariant mass and rapidity separation of the 
leading and sub-leading jets, \emph{i.e.}\ on the two jets with largest
transverse momenta, the following VBS cuts are applied:
\begin{align}
\label{eq:vbscuts}
 m_{\Pj \Pj} >  500\GeV,\qquad |\Delta y_{\Pj \Pj}| > 2.5.
\end{align}
Note that the two leading jets are used in the definition of the
dynamical scale in Eq.~\eqref{eq:defscale} and are also referred to as
tagging jets.

\subsection{Integrated cross section}
\label{ssec:IntegratedCrossSection}

We start by reporting the fiducial cross section at leading order in \refta{table:LOcrosssection}.
\begin{table}
\begin{center}
\begin{tabular}{|l||c|c|c||c|}
\hline
Order & $\mathcal{O}{\left(\alpha^{6}\right)}$ & $\mathcal{O}{\left(\alphas\alpha^{5}\right)}$ & $\mathcal{O}{\left(\alphas^2\alpha^{4}\right)}$ & Sum \\
\hline
\hline
${\sigma_{\mathrm{LO}}}$ [fb] 
& $1.4178(2)$
& $0.04815(2)$
& $0.17229(5)$
& $1.6383(2)$ \\
\hline
\hline
$\sigma^{\rm max}_{\mathrm{LO}}$ [fb] 
& $1.5443(2)$
& $0.05680(3)$
& $0.22821(6)$
& $1.8293(2)$ \\
\hline
$\sigma^{\rm min}_{\mathrm{LO}}$ [fb] 
& $1.3091(2)$
& $0.04135(2)$
& $0.13323(3)$
& $1.4836(2)$ \\
\hline
\end{tabular}
\end{center}
\caption{
Fiducial cross section at LO for the process
$\Pp\Pp\to\mu^+\nu_\mu\Pe^+\nu_{\Pe}\Pj\Pj$, stated separately for the
orders  
$\mathcal{O}{\left(\alpha^{6}\right)}$, $\mathcal{O}{\left(\alpha_{\rm
      s}\alpha^{5}\right)}$, and $\mathcal{O}{\left(\alpha_{\rm
      s}^2\alpha^{4}\right)}$ and for the sum of all the LO
contributions expressed in femtobarn. 
The cross section ${\sigma_{\mathrm{LO}}}$ corresponds to the central
scale choice,
while the cross sections $\sigma^{\rm max}_{\mathrm{LO}}$ and $\sigma^{\rm min}_{\mathrm{LO}}$ correspond to the scale choices leading to the maximum and minimum cross section, respectively.
The statistical uncertainty from the Monte Carlo integration on the last digit is given in parenthesis.}
\label{table:LOcrosssection}
\end{table}
The scale dependence of the results has been studied upon varying the
factorisation and renormalisation scales independently. Specifically,
the central scale defined in Eq.~\eqref{eq:defscale} has been scaled by factors $\xi_{\rm fac}$ and $\xi_{\rm ren}$ for the combinations
\begin{align}
\label{eq:scales}
 (\xi_{\rm fac},\xi_{\rm ren}) \in \big\{\left(1/2,1/2\right),\,
 \left(1/2,1\right),\,
 \left(1,1/2\right),\,
 \left(1,1\right),\,
 \left(1,2\right),\,
 \left(2,1\right),\,
 \left(2,2\right)\big\},
\end{align}
where $(\xi_{\rm fac},\xi_{\rm ren}) = \left(1,1\right)$ corresponds
to the central scale.  For each cross section, three values are given:
the one corresponding to the central scale, the maximum, and the
minimum.  For the fiducial cross section, the sum of the contributions
of all orders is computed for each scale choice separately, and then
the maximum and the minimum are extracted.  The order
$\mathcal{O}{\left(\alpha^{6}\right)}$ corresponds to the EW-induced
contribution, the order $\mathcal{O}{\left(\alpha_{\rm
      s}^2\alpha^{4}\right)}$ to the QCD-induced contribution, and the
order $\mathcal{O}{\left(\alphas\alpha^{5}\right)}$ represents the
interferences.  For the fiducial volume with VBS cuts defined in the
previous section, the EW-induced process is clearly dominating over
its irreducible background processes.  It amounts to $87\%$ of the
cross section of the full process
$\Pp\Pp\to\mu^+\nu_\mu\Pe^+\nu_{\Pe}\Pj\Pj$, while the
$\mathcal{O}{\left(\alphas^2\alpha^{4}\right)}$ contributions add up
to about $10\%$.  The impact of the interferences on the fiducial
cross section is small, at the level of $3\%$.  The contribution of
individual channels is actually larger since interferences enter with
positive and negative sign (\emph{e.g.}\ $+4\%$ for the $\Pu\Pu$
channel and $-1.2\%$ for the $\Pu\bar\Pd$ channels) 
and not all
channels involve interferences.  The smallness of the interferences is
not unexpected, since by construction, resonances in interfered
$t$--$u$-channel or $s$--$t/u$-channel diagrams are suppressed with
respect to kinematic topologies from squared resonant $s$-, $t$- or
$u$-channel diagrams present in the order
$\mathcal{O}{\left(\alpha^{6}\right)}$ and
$\mathcal{O}{\left(\alphas^2\alpha^{4}\right)}$ contributions.  At
leading order, we find a scale dependence of
$\left[+8.9\%;-7.7\%\right]$, $\left[+17.9\%;-14.1\%\right]$,
$\left[+32.5\%;-22.7\%\right]$ for the contributions of orders
$\mathcal{O}{\left(\alpha^{6}\right)}$, $\mathcal{O}{\left(\alpha_{\rm
      s}^1\alpha^{5}\right)}$, $\mathcal{O}{\left(\alpha_{\rm
      s}^2\alpha^{4}\right)}$, respectively, leading to
 \begin{equation}
\sigma_{\rm LO} = 1.6383(2)_{-9.44(2)\%}^{+11.66(2)\%}\fb.
\end{equation}
Naturally the scale dependence is larger for contributions
depending on the strong coupling.

In \refta{table:NLOcrosssection}, all NLO corrections to the
fiducial cross sections split into contributions of the 
different orders in the strong and EW coupling are presented. 
\begin{table}
\begin{center}
\begin{tabular}{|l||c|c|c|c||c|}
\hline
Order & $\mathcal{O}{\left(\alpha^{7}\right)}$ & $\mathcal{O}{\left(\alphas\alpha^{6}\right)}$ & $\mathcal{O}{\left(\alphas^{2}\alpha^{5}\right)}$ & $\mathcal{O}{\left(\alphas^{3}\alpha^{4}\right)}$ & Sum \\
\hline
\hline 
${\delta \sigma_{\mathrm{NLO}}}$ [fb] 
& $-0.2169(3)$ 
& $-0.0568(5)$
& $-0.00032(13)$
& $-0.0063(4)$ 
& $-0.2804(7)$ \\
\hline
$\delta \sigma_{\mathrm{NLO}}/\sigma_{\rm LO}$ [\%] & $-13.2$ & $-3.5$ & $0.0$ & $-0.4$ & $-17.1$ \\
\hline
\end{tabular}
\end{center}
\caption{
NLO corrections for the process $\Pp\Pp\to\mu^+\nu_\mu\Pe^+\nu_{\Pe}\Pj\Pj$ at the orders 
$\mathcal{O}{\left(\alpha^{7}\right)}$, $\mathcal{O}{\left(\alphas\alpha^{6}\right)}$, $\mathcal{O}{\left(\alphas^{2}\alpha^{5}\right)}$, and $\mathcal{O}{\left(\alphas^{3}\alpha^{4}\right)}$ and for the sum of all NLO corrections.
The contribution $\delta\sigma_{\mathrm{NLO}}$ corresponds to the absolute correction for the central scale choice while $\delta \sigma_{\mathrm{NLO}}/\sigma_{\rm LO}$ gives the relative correction normalised to the sum of all LO contributions at the central scale.
The absolute contributions are expressed in femtobarn while the relative ones are expressed in per cent.
The statistical uncertainty from the Monte Carlo integration on the last digit is given in parenthesis.}
\label{table:NLOcrosssection}
\end{table}
In the following, the relative NLO corrections are always normalised
to the sum of all LO contributions.  The total correction to the full
process is large and negative, amounting to $-17.1\%$.  
The bulk
of the correction with $-13.2\%$ stems from the order
$\mathcal{O}{\left(\alpha^{7}\right)}$, the EW correction to the EW-induced
process.  Note that the correction is smaller than the $-16.0\%$
stated in \citere{Biedermann:2016yds}, mainly owing to the
normalisation to the sum of all LO contributions instead of to the
$\mathcal{O}{\left(\alpha^{6}\right)}$ contribution alone. The
remaining additional difference due to the dynamical scale choice is
small ($+0.7\%$) as this affects the purely EW contribution only via
the evolution of the PDF and not via the running in $\alphas$.
The second-largest corrections with $-3.5\%$ occur at order
$\mathcal{O}{\left(\alphas\alpha^{6}\right)}$.  The
contribution of order  $\mathcal{O}{\left(\alphas^{3}\alpha^{4}\right)}$ with a correction of $-0.4\%$ is suppressed by another order of magnitude.
The contribution of order $\mathcal{O}{\left(\alphas^{2}\alpha^{5}\right)}$ is even more suppressed and phenomenologically unimportant at the fiducial cross-section level.
The hierarchy of the NLO corrections follows roughly the pattern
observed at LO: at the integrated cross-section level, each NLO
correction is roughly one order of
magnitude smaller than the corresponding LO contribution.  Thus, one expects that the
bulk of the $\mathcal{O}{\left(\alphas\alpha^{6}\right)}$
corrections stems from the QCD corrections to the EW-induced process,
while only a small contribution results from the EW corrections to
the interference.
We emphasise, however, again that QCD corrections to the EW-induced process
and EW corrections to the LO interference cannot be defined
independently.  Indeed, using the full matrix element, they both
contribute at the order $\mathcal{O}{\left(\alpha_{\rm
      s}\alpha^{6}\right)}$ as discussed in \refse{ssec:VirtualCorrections}.
The contributions at the order $\mathcal{O}{\left(\alpha_{\rm
      s}^{2}\alpha^{5}\right)}$ are small because the corresponding LO
contributions are already suppressed and moreover the EW corrections
to the QCD-induced LO contribution and the QCD corrections to the LO
interference cancel to a large extent.
Upon calculating the NLO cross section with the different scales of
Eq.~\eqref{eq:scales}, we find
\begin{equation} 
\sigma_{\rm NLO} =  1.3577(7)_{-2.7(1)\%}^{+1.2(1)\%}\fb, 
\end{equation}
\emph{i.e.}\ a reduction of the LO scale dependence by a factor {five}.

We have also calculated the photon-induced NLO contributions as shown
in \refta{table:photoninducedXsection}. Since the photon PDF from the
NNPDF-3.0 QED set is known to give rather sizeable 
contributions with a large error,
we have also calculated these contributions using the PDF of
the recent 
$\mathrm{LUXqed}\_\mathrm{plus}\_\mathrm{PDF4LHC15}\_\mathrm{nnlo}\_100$
set~\cite{Manohar:2016nzj}. For LUXqed we use the ${\overline{\rm MS}}$
factorisation scheme throughout, while 
we have verified that 
the effect of the factorisation
scheme is irrelevant at the level of accuracy of the results given.
The photon-induced NLO contributions are dominated by those of order
$\mathcal{O}{\left(\alpha^{7}\right)}$ and amount to $2.7\%$ based on
NNPDF-3.0 QED and $1.5\%$ based on LUXqed.  The photon-induced
contributions of orders $\mathcal{O}{\left(\alphas\alpha^{6}\right)}$
and $\mathcal{O}{\left(\alphas^2\alpha^{5}\right)}$ are negligible.
Hence in the following, only the photon-induced contributions of order $\mathcal{O}{\left(\alpha^{7}\right)}$ are displayed in the distributions.
Note that in our definition of the NLO corrections at order
$\mathcal{O}{\left(\alpha^{7}\right)}$, the photon-induced
contributions are not included 
but are shown separately.
This means that for the combined distributions (Fig.~\ref{fig:distributions_scale}), the NLO predictions do not include the photon-induced contributions.
\begin{table}
\begin{center}
\begin{tabular}{|l|l||c|c|c|}
\hline
Order & PDF & $\mathcal{O}{\left(\alpha^{7}\right)}$ & $\mathcal{O}{\left(\alphas\alpha^{6}\right)}$ & $\mathcal{O}{\left(\alphas^{2}\alpha^{5}\right)}$ \\
\hline
\hline
${\delta \sigma_{\mathrm{NLO}}}$ [fb] & NNPDF-3.0 QED
& $0.04368(2)$ 
& $<10^{-6}$ 
& $0.000074(1)$ \\ 
\hline
$\delta \sigma_{\mathrm{NLO}}/\sigma_{\rm LO}$ [\%] & NNPDF-3.0 QED & $+2.66$ & $<0.0001$ & $+0.004$ \\
\hline
\hline
$\delta \sigma_{\mathrm{NLO}}/\sigma_{\rm LO}$ [\%] & LUXqed & $+1.51$ & $<0.0001$ & $+0.002$ \\
\hline
\end{tabular}
\end{center}
\caption{
Photon-induced NLO corrections for the process $\Pp\Pp\to\mu^+\nu_\mu\Pe^+\nu_{\Pe}\Pj\Pj$ at the orders 
$\mathcal{O}{\left(\alpha^{7}\right)}$, $\mathcal{O}{\left(\alpha_{\rm
      s}\alpha^{6}\right)}$, and $\mathcal{O}{\left(\alpha_{\rm
      s}^{2}\alpha^{5}\right)}$ in both absolute (expressed in femtobarn) and relative value (expressed in per cent) for the PDF set NNPDF-3.0 QED. In addition, the relative
corrections are also given for the LUXqed PDF set.}
\label{table:photoninducedXsection}
\end{table}

So far, all computations in the literature 
at the order $\mathcal{O}{\left(\alphas\alpha^{6}\right)}$
\cite{Jager:2009xx,Jager:2011ms,Denner:2012dz} have been done in the
so-called VBS approximation.  This features the inclusion of $t$- and
$u$-channel diagrams but neglects their interferences as well as
$s$-channel contributions.  In order to assess the quality of this
approximation, we have re-computed the full
$\mathcal{O}{\left(\alphas\alpha^{6}\right)}$ corrections without VBS approximation in the
set-ups of \citeres{Jager:2009xx} and \cite{Denner:2012dz}.  
In \refta{table:comparison1}, a comparison of LO
and NLO fiducial cross sections is presented 
in the set-up of \citere{Jager:2009xx}.  In \refta{table:comparison2}, 
results are compared with those of
\citere{Denner:2012dz}\footnote{Note that the LO cross section
  reported here for \citere{Denner:2012dz} corresponds to the
  approximate calculation.}.  At the level of the
fiducial cross section,
 the approximate calculations turn out to agree 
within $0.6\%$ with the full computation presented 
here at both LO [order $\mathcal{O}{\left(\alpha^{6}\right)}$] and NLO
[order $\mathcal{O}{\left(\alphas\alpha^{6}\right)}$]. In addition, a modified version of \recola allowed us to confirm a difference of $0.6\%$ at NLO between the full computation and the VBS approximation in our set-up.

\begin{table}
\begin{center}
\begin{tabular}{|l||c|c|c|}
\hline
 Set-up of \citere{Jager:2009xx} & Present work & DHK~\cite{Denner:2012dz} & JOZ~\cite{Jager:2009xx} \\
\hline
\hline
${\sigma_{\mathrm{LO}}}$ [fb] 
& $1.4038(4)$
& $1.4061(7)$
& $1.409$ \\
\hline
\hline
$\sigma_{\mathrm{NLO}}$ [fb] 
& $1.380(1)$
& $1.372(1)$
& $1.372$ \\
\hline
\end{tabular}
\end{center}
\caption{
Comparison of fiducial cross sections at LO [order
$\mathcal{O}{\left(\alpha^{6}\right)}$] and NLO [order
$\mathcal{O}{\left(\alphas\alpha^{6}\right)}$] for the process 
$\Pp\Pp\to\mu^+\nu_\mu\Pe^+\nu_{\Pe}\Pj\Pj$ against the literature 
in the set-up of \citere{Jager:2009xx} with MSTW08 PDF.
DHK denotes the calculation of \citere{Denner:2012dz}, while JOZ
refers to the one of \citere{Jager:2009xx}. 
The cross sections are expressed in femtobarn and the statistical uncertainty from the Monte Carlo integration on the last digit is given in parenthesis.}
\label{table:comparison1}
\end{table}

\begin{table}
\begin{center}
\begin{tabular}{|l||c|c|}
\hline
 Set-up of \citere{Denner:2012dz} & Present work & DHK~\cite{Denner:2012dz} \\
\hline
\hline
${\sigma_{\mathrm{LO}}}$ [fb] 
& $1.2230(4)$
& $1.2218(2)$ \\
\hline
\hline
$\sigma_{\mathrm{NLO}}$ [fb] 
& $1.2975(15)$
& $1.2917(8)$ \\
\hline
\end{tabular}
\end{center}
\caption{
Comparison of fiducial cross sections at LO [order
$\mathcal{O}{\left(\alpha^{6}\right)}$] and NLO [order
$\mathcal{O}{\left(\alphas\alpha^{4}\right)}$] for the process 
$\Pp\Pp\to\mu^+\nu_\mu\Pe^+\nu_{\Pe}\Pj\Pj$ against the literature
in the set-up of \citere{Denner:2012dz}.
DHK denotes the results of \citere{Denner:2012dz}.
The cross sections are expressed in femtobarn and the statistical uncertainty from the Monte Carlo integration on the last digit is given in parenthesis.}
\label{table:comparison2}
\end{table}

\subsection{Differential distributions}
\label{ssec:DifferentialDistributions}

We start the discussion of differential distributions with plots
showing all the different contributions in the strong and EW coupling
at both LO and NLO.  In the upper panel, the three LO contributions as
well as the full NLO prediction are plotted.  In the two lower panels,
the four contributions to the relative NLO corrections normalised to
the sum of all the LO contributions are presented along with the NLO
photon-induced contributions of order
$\mathcal{O}{\left(\alpha^{7}\right)}$.  The latter are computed for
the LUXqed PDF and are thus normalised to the Born contributions
obtained with the corresponding PDF.  Remember that these
photon-induced contributions are not included in our definition of the
NLO corrections of order $\mathcal{O}{\left(\alpha^{7}\right)}$.

\begin{figure}
        \setlength{\parskip}{-10pt}
        \begin{subfigure}{0.49\textwidth}
                \subcaption{}
                \includegraphics[width=\textwidth]{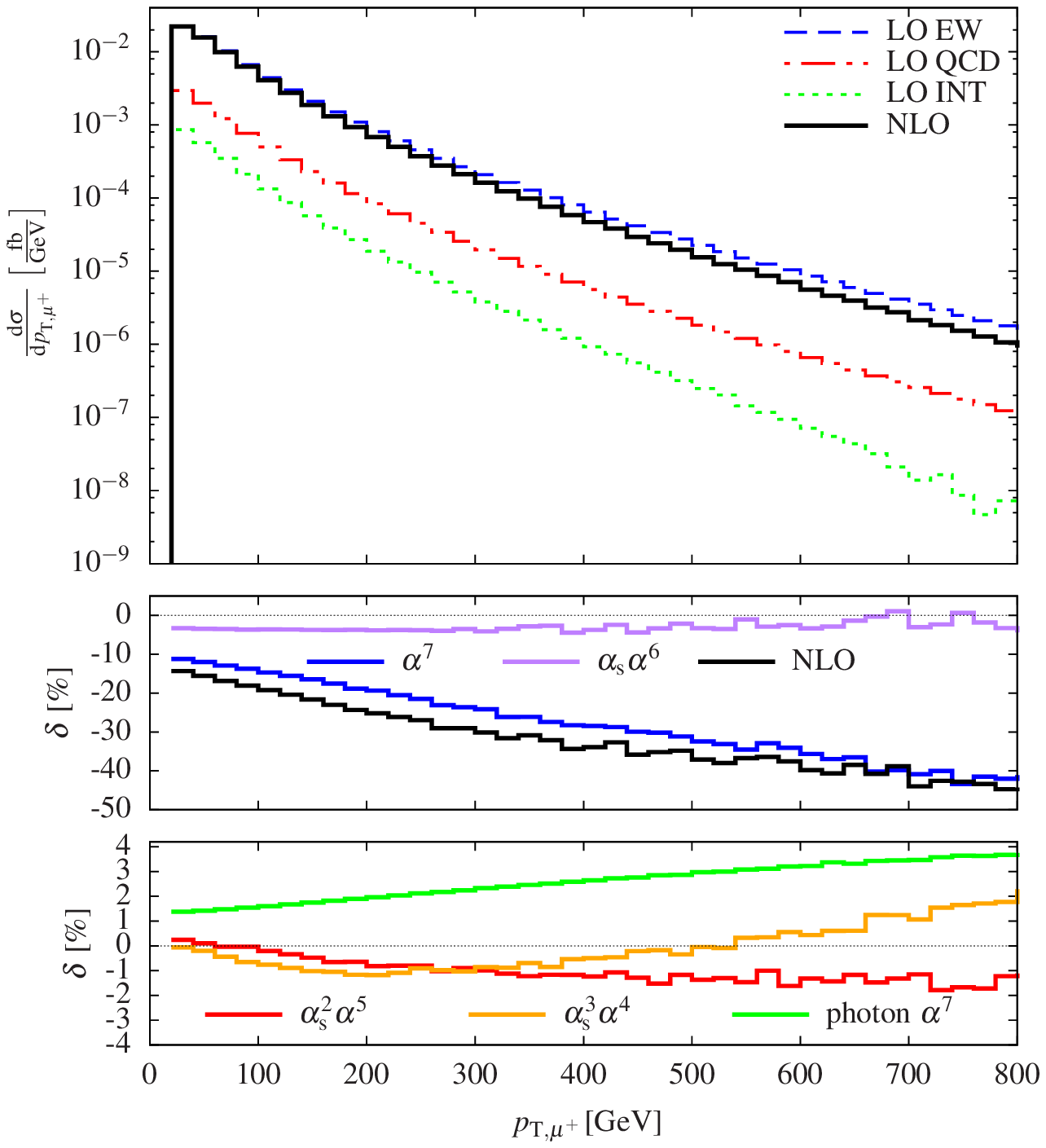}
                \label{plot:transverse_momentum_antimuon}
        \end{subfigure}
        \hfill
        \begin{subfigure}{0.49\textwidth}
                \subcaption{}
                \includegraphics[width=\textwidth]{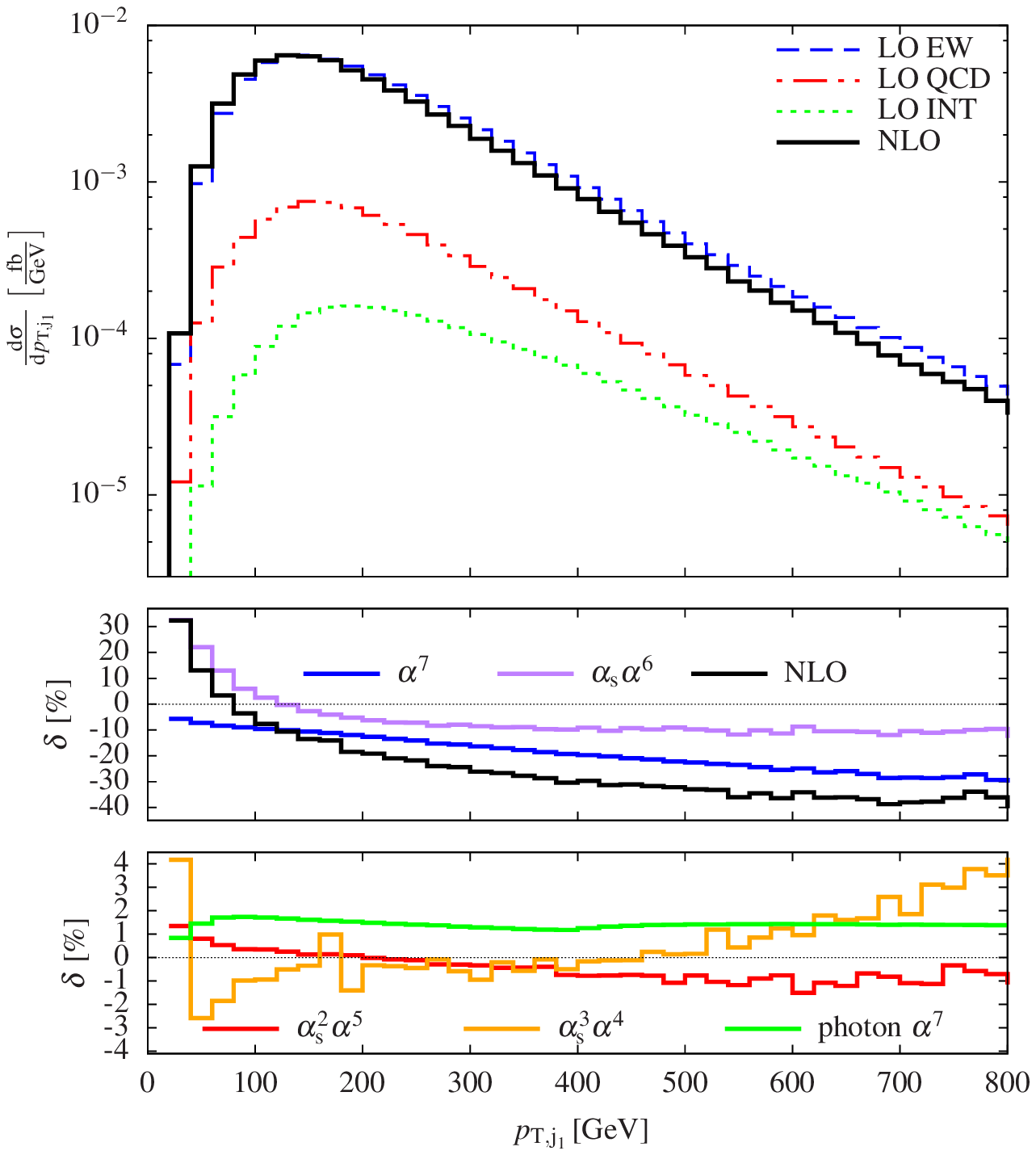}
                \label{plot:transverse_momentum_harder_jet} 
        \end{subfigure}
     
        \vspace*{-3ex}
        \caption{\label{fig:transverse_momentum_distributions}%
                Transverse-momentum distributions at a centre-of-mass energy $\sqrt{s}=13\TeV$ at the LHC for $\Pp\Pp\to\mu^+\nu_\mu\Pe^+\nu_{\Pe}\Pj\Pj$: 
                \subref{plot:transverse_momentum_antimuon} for the anti-muon~(left) 
                and %
                \subref{plot:transverse_momentum_harder_jet} the hardest jet~(right).
                The upper panels show the three LO contributions as well as the sum of all NLO predictions.
                The two lower panels show the relative NLO corrections with respect to the full LO, defined as $\delta_i = \delta \sigma_{i} / \sum \sigma_{\text{LO}}$, 
                where $i=\mathcal{O}{\left(\alpha^{7}\right)},\mathcal{O}{\left(\alphas\alpha^{6}\right)},\mathcal{O}{\left(\alphas^2\alpha^{5}\right)},\mathcal{O}{\left(\alphas^3\alpha^{4}\right)}$.
                In addition, the NLO photon-induced contributions of order $\mathcal{O}{\left(\alpha^{7}\right)}$ computed with LUXqed is provided separately.}
\end{figure}
In \reffi{fig:transverse_momentum_distributions}, two
transverse-momentum distributions are displayed.
Starting with the distribution in the  transverse 
momentum of the anti-muon, the upper panel  in
\reffi{plot:transverse_momentum_antimuon} shows  that the EW-induced 
contribution is dominant over the whole phase space. 
Concerning the relative NLO corrections in the lower panel, the largest 
contribution is the one of order $\mathcal{O}{\left(\alpha^{7}\right)}$.
It ranges from $-10\%$ at $20\GeV$ (the cut on the transverse momentum
of the charged lepton) to $-40\%$ at $800\GeV.$ The large corrections
for high transverse momenta are due to logarithms of EW origin, the
so-called Sudakov logarithms, as already pointed out in
\citere{Biedermann:2016yds}.  The second largest contribution is the
one of order $\mathcal{O}{\left(\alphas\alpha^{6}\right)}$ which
consists of QCD corrections to the EW-induced contribution and EW
corrections to the interference.  
Over the considered range, this contribution stays between $-4\%$ and
$0\%$.  While the corrections of order
$\mathcal{O}{\left(\alphas^{2}\alpha^{5}\right)}$ are negligible at
the level of the fiducial cross section, they reach $-2\%$ for
$p_{\rT,\mu^-}=800\GeV$.  The corrections of orders
$\mathcal{O}{\left(\alphas^{3}\alpha^{4}\right)}$ stay also below
$2\%$ in magnitude and cancel those of order
$\mathcal{O}{\left(\alphas^{2}\alpha^{5}\right)}$ for large
$p_{\rT,\mu^-}$.
While the $\mathcal{O}{\left(\alphas^{2}\alpha^{5}\right)}$
contributions decrease owing to the presence of Sudakov logarithms,
the $\mathcal{O}{\left(\alphas^{3}\alpha^{4}\right)}$ contributions
steadily increase above $p_{\rT,\mu^+}=200\GeV$.  The photon-induced
contributions increase from $1.5\%$ to $4\%$ with increasing
$p_{\rT,\mu^+}$, while for other distributions they are smaller and
mostly do not show any shape distortion.  This is in accordance with
what has been found for LO photon-induced contributions for $\Pp \Pp
\to \Pe^+ \nu_{\Pe} \mu^- \bar{\nu}_\mu \Pb \bar{\Pb} \PH$
\cite{Denner:2016wet}.

Figure \ref{plot:transverse_momentum_harder_jet} shows the
distribution in the transverse momentum of the leading jet.  While
there was a clear hierarchy between the LO contributions in the
previous observable, here the LO interference becomes comparable to
the QCD-induced process around $800\GeV$ (see also
\citere{Campanario:2013gea}).  Strikingly, the shape of the
distribution at low transverse momentum is rather different.  By
construction, for small transverse momentum of the leading jet, the
transverse momenta of all sub-leading jets must be small as well. This
suppresses the available phase space and explains why the LO
distribution is less peaked at small transverse momenta as compared to
the distributions in the transverse momentum of the anti-muon
(\reffi{plot:transverse_momentum_antimuon}) or the transverse momentum
of the sub-leading jet (not shown).  Concerning the NLO contributions,
the main difference with respect to the distribution in the transverse
momentum of the muon is the behaviour of the
$\mathcal{O}{\left(\alphas\alpha^{6}\right)}$ corrections: they are
large and positive at the kinematical threshold of $30\GeV$ (at the
level of $30\%$), decrease towards $-5\%$ around $200\GeV$ and stay
almost constant over the whole spectrum up to $800\GeV$.
The large QCD corrections for small $p_{\rT,\Pj_1}$ have 
already been observed in \citere{Denner:2012dz} for the
same dynamical scale.  The strong increase of the QCD corrections at
low $p_{\rm T}$, which is also present in the results of
\citere{Campanario:2013gea}, is a kinematical effect genuine to the
distribution in the transverse momentum of the leading jet and is
independent of the QCD scale.  As observed also in
Refs.~\cite{Cacciari:2015jma,Rauch:2017cfu} at NLO and NNLO QCD for
Higgs-boson production in vector-boson fusion, the QCD corrections
have the effect of redistributing jets from higher to lower transverse
momenta.  This behaviour is mainly driven by the real radiation and
causes a large effect for small $p_{\rT,j_1}$ where the LO
contribution is suppressed.
The contributions  of orders
$\mathcal{O}{\left(\alphas^{2}\alpha^{5}\right)}$ and
$\mathcal{O}{\left(\alphas^{3}\alpha^{4}\right)}$ behave qualitatively
similar as for the $p_{\rT,\mu^+}$ distribution.

\begin{figure}
        \setlength{\parskip}{-10pt}
        \begin{subfigure}{0.49\textwidth}
                \subcaption{}
                \includegraphics[width=\textwidth]{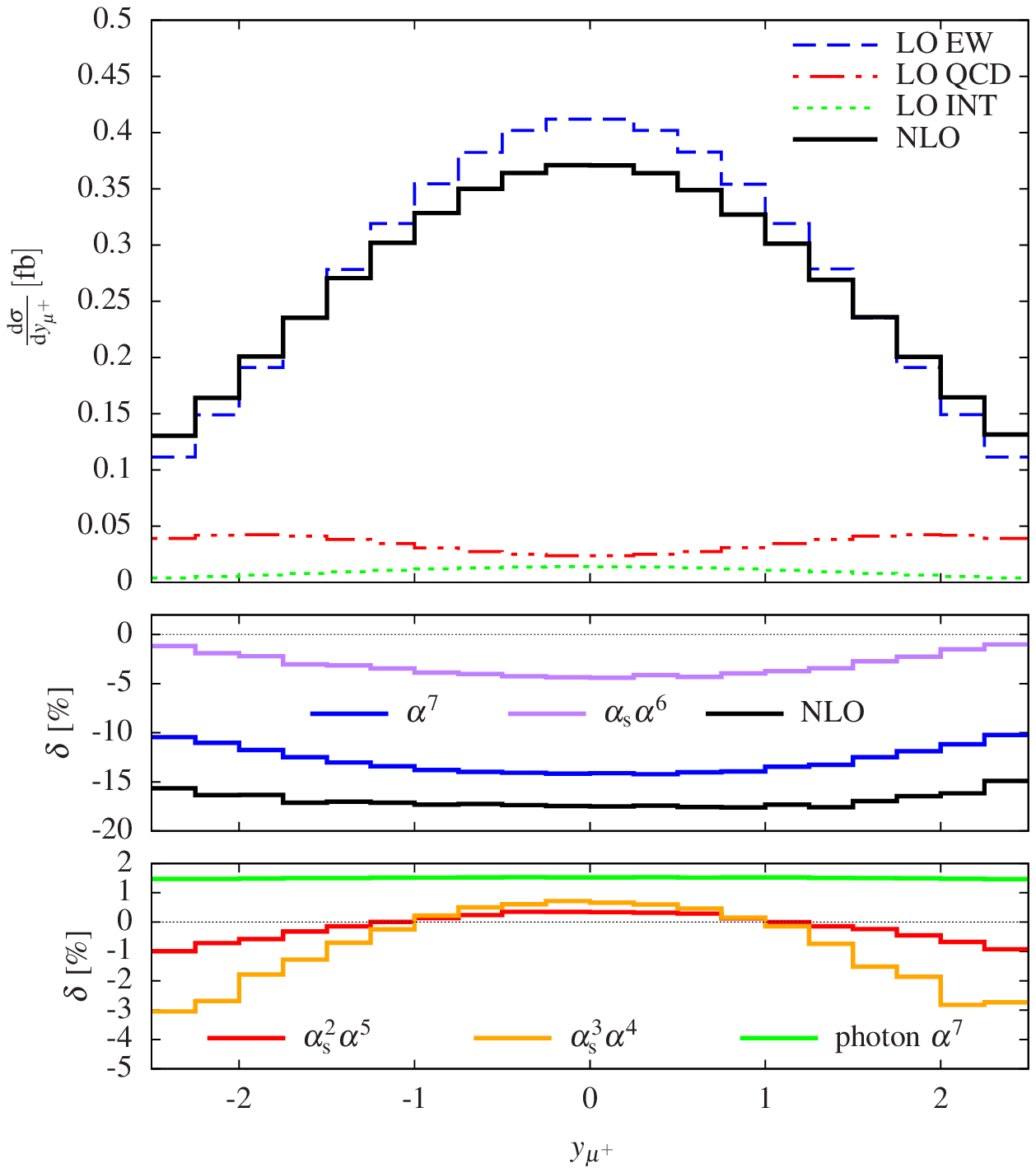}
                \label{plot:rapidity_antimuon}
        \end{subfigure}
        \hfill
        \begin{subfigure}{0.49\textwidth}
                \subcaption{}
                \includegraphics[width=\textwidth]{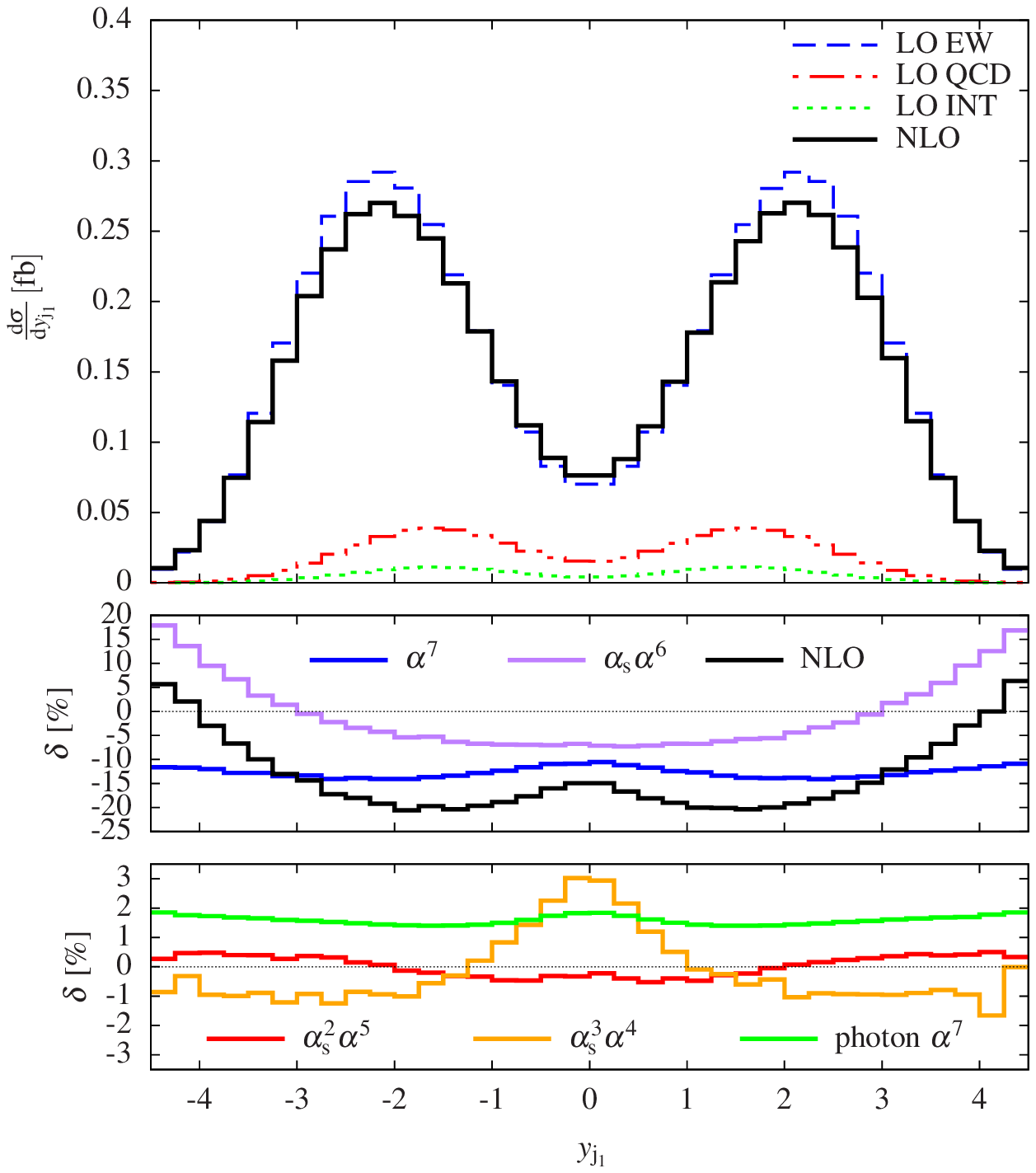}
                \label{plot:rapidity_j1} 
        \end{subfigure}
        
        \begin{subfigure}{0.49\textwidth}
                \subcaption{}
                \includegraphics[width=\textwidth]{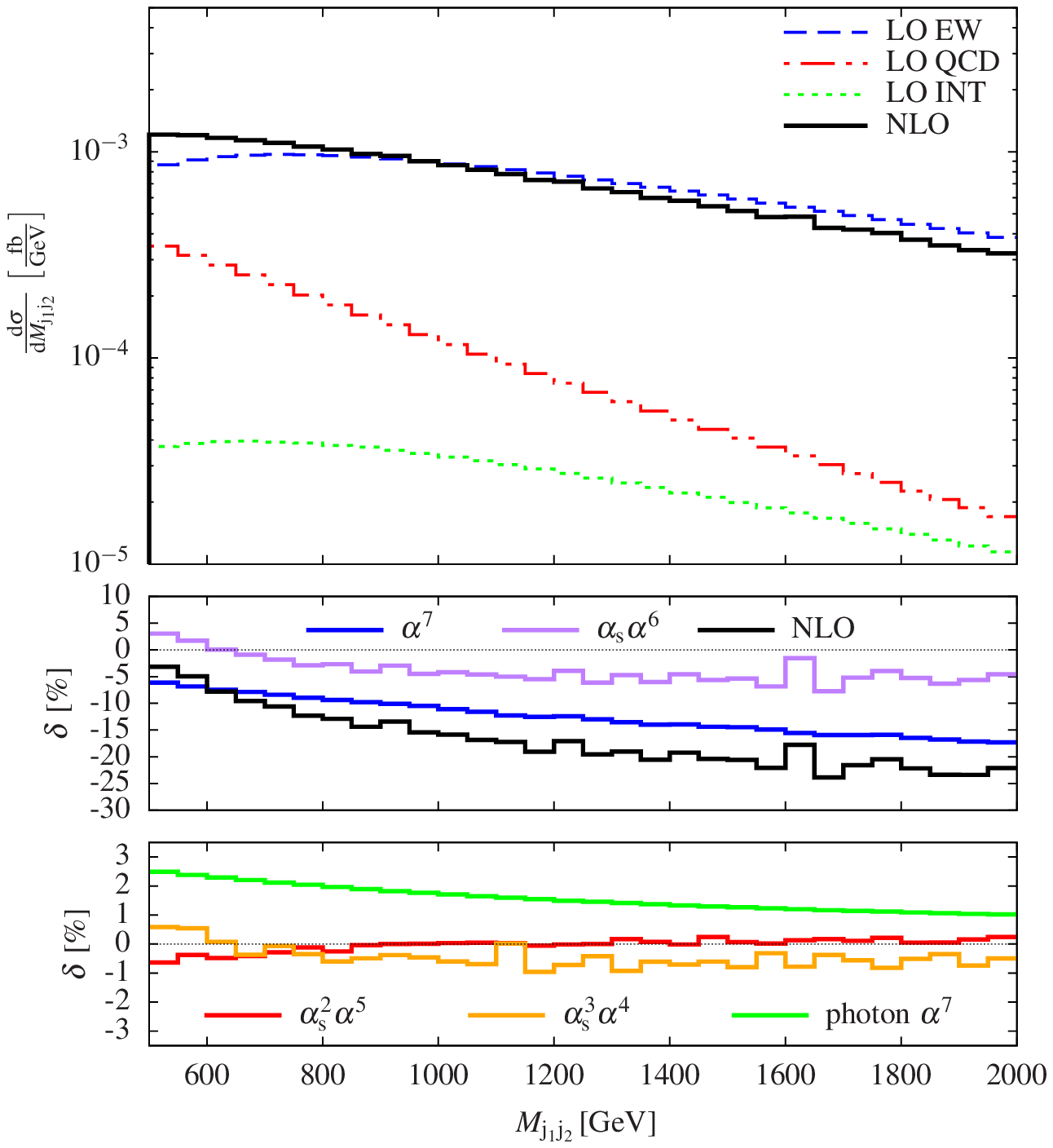}
                \label{plot:invariant_mass_mjj12}
        \end{subfigure}
        \hfill
        \begin{subfigure}{0.49\textwidth}
                \subcaption{}
                \includegraphics[width=\textwidth]{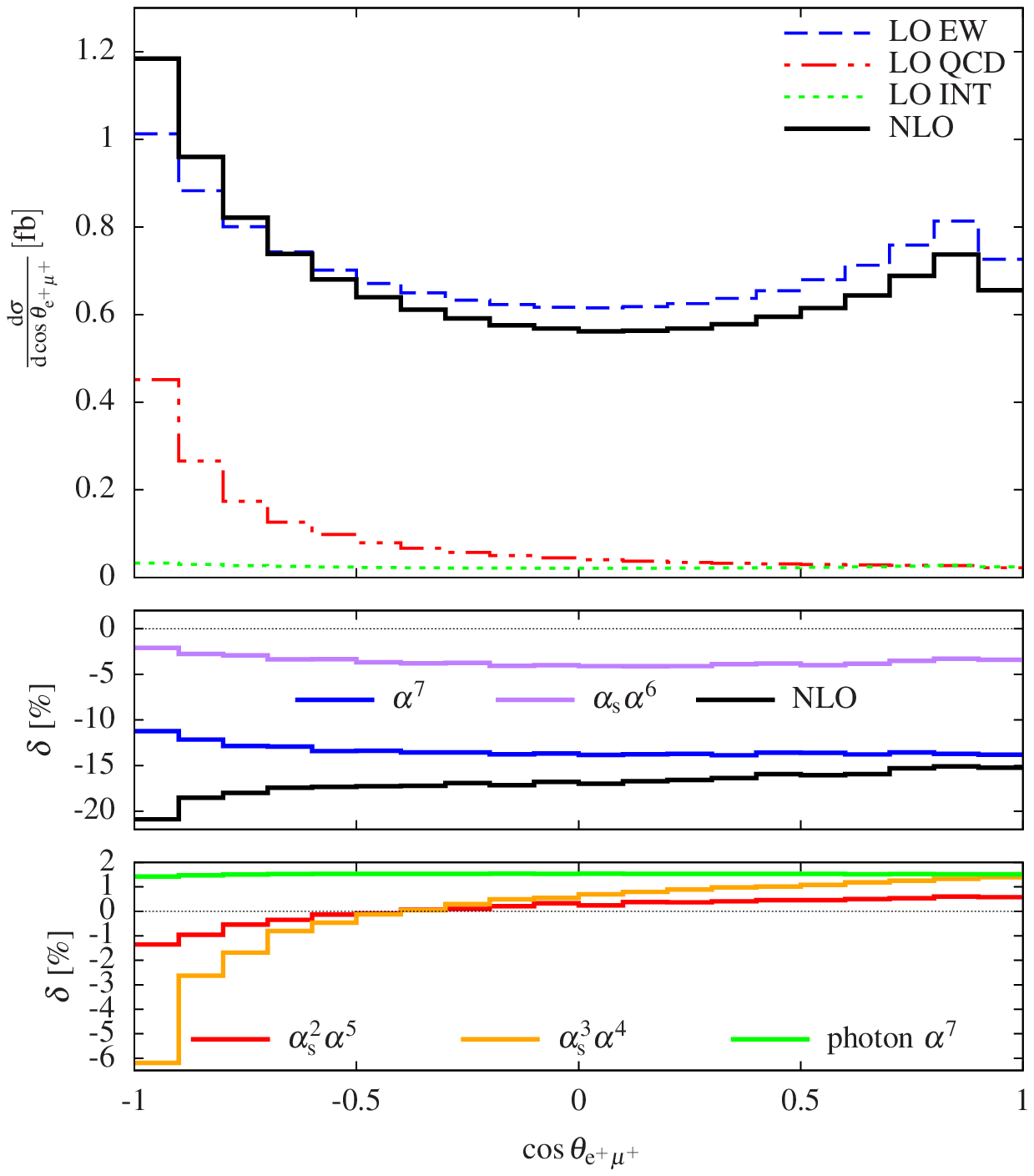}
                \label{plot:cosine_angle_separation_epmu}
        \end{subfigure}
        
        \vspace*{-3ex}
        \caption{\label{fig:various_distributions}%
                Differential distributions at a centre-of-mass energy $\sqrt{s}=13\TeV$ at the LHC for $\Pp\Pp\to\mu^+\nu_\mu\Pe^+\nu_{\Pe}\Pj\Pj$: 
                \subref{plot:rapidity_antimuon}~rapidity for the anti-muon~(top left), %
                \subref{plot:rapidity_j1}~rapidity for the hardest jet~(top right),
                \subref{plot:invariant_mass_mjj12}~invariant mass for the two leading jets~(bottom left), and
                \subref{plot:cosine_angle_separation_epmu}~cosine of
                the angle between the positron and the anti-muon~(bottom right).
                The upper panels show the three LO contributions as well as the sum of all NLO predictions.
                The two lower panels show the relative NLO corrections with respect to the full LO in per cent,
                defined as $\delta_i = \delta \sigma_{i} / \sum \sigma_{\text{LO}}$, 
                where $i=\mathcal{O}{\left(\alpha^{7}\right)},\mathcal{O}{\left(\alphas\alpha^{6}\right)},\mathcal{O}{\left(\alphas^2\alpha^{5}\right)},\mathcal{O}{\left(\alphas^3\alpha^{4}\right)}$.
                In addition, the NLO photon-induced contributions of order $\mathcal{O}{\left(\alpha^{7}\right)}$ computed with LUXqed is provided separately.}
\end{figure}
The distributions in the rapidities of the anti-muon and of the hardest
jet are displayed in \reffis{plot:rapidity_antimuon} and
\ref{plot:rapidity_j1}, respectively.  As
\reffi{plot:rapidity_antimuon} shows, the anti-muons are mostly
produced in the central region of the detector for the EW-induced process,
while for the QCD-induced contribution, although suppressed, they lie
preferentially at rapidities around $\pm2$. The interference, even
more suppressed, is largest in the central region.  The relative NLO
contributions also display different behaviours.  The contributions of
order $\mathcal{O}{\left(\alpha^{7}\right)}$ and
$\mathcal{O}{\left(\alphas\alpha^{6}\right)}$ are maximally
negative in the central region and decrease 
in magnitude
in the peripheral region.
On the other hand, the contributions of order
$\mathcal{O}{\left(\alphas^{2}\alpha^{5}\right)}$ and
$\mathcal{O}{\left(\alphas^{3}\alpha^{4}\right)}$ 
display an opposite behaviour with a 
small positive maximum in the central region and larger negative
corrections in the forward and backward directions, which is mainly
caused by the increased relative size of the QCD-induced LO
contributions.  
Like for the fiducial cross section, the hierarchy of
the corrections follows closely the LO contributions which is expected
as the anti-muon rapidity distribution is rather flat.

The rapidity of the hardest jet (\reffi{plot:rapidity_j1}) also
displays interesting patterns.  In the absolute prediction, one sees
the typical VBS kinematic with central rapidity gap.
The EW corrections to the EW-induced contribution of order
$\mathcal{O}{\left(\alpha^{7}\right)}$ are larger in the region where
the VBS process is dominating while they are smaller in the central
region, which is typically dominated by non-VBS configurations.  This
observable has already been discussed in \citere{Biedermann:2016yds}.
The corrections of order
  $\mathcal{O}{\left(\alphas\alpha^{6}\right)}$ are negative
  in the central region at a level of $-5\%$, but around
  $+18\%$ for rapidity of $\pm4.5$.  
Such a behaviour of the QCD
corrections was also found in \citere{Denner:2012dz} for the
considered process and for Higgs production via vector-boson fusion in
\citere{Figy:2003nv}.  Hence the QCD corrections have the effect of
making the leading jet more forward.
The corrections of order
$\mathcal{O}{\left(\alphas^{2}\alpha^{5}\right)}$ 
are flat and below $1\%$ in magnitude. Those of order
$\mathcal{O}{\left(\alphas^{3}\alpha^{4}\right)}$ 
reach $+3\%$ in the central region but are at the level of $-1\%$ for
large rapidities. 

In \reffi{plot:invariant_mass_mjj12}, the distribution in the
invariant mass of the two tagging jets is displayed.  As pointed out
already in \citere{Biedermann:2016yds}, at LO the VBS contribution
extends to large invariant masses. 
The QCD-induced one drops significantly faster to become an order of
magnitude smaller than the VBS contribution at $1200\GeV$.  This
illustrates the need for VBS-specific event selections.  Indeed, by
extrapolating to lower invariant mass, it is clear that in this region
the QCD-induced process would be sizeable.  As for the EW-induced
process, the interference contribution displays a comparably flat
behaviour becoming of the same size as the QCD-induced one around
$2000\GeV$.  The relative NLO corrections are similar to those for the
distributions in the transverse momenta.
The EW corrections to the EW-induced process display the typical
behaviour of Sudakov logarithms in the high-invariant-mass region and
grow negatively large towards high invariant masses.  The contributions
of order $\mathcal{O}{\left(\alphas\alpha^{6}\right)}$ are positive
for $M_{\Pj_1\Pj_2}=500\GeV$ but tend to $-5\%$ at high invariant
masses.  The contributions of order
$\mathcal{O}{\left(\alphas^{2}\alpha^{5}\right)}$ and
$\mathcal{O}{\left(\alphas^{3}\alpha^{4}\right)}$ are below $1\%$ in
magnitude and tend to compensate each other.
The photon-induced contributions slightly exceed $2\%$ for small
invariant mass and decrease for higher invariant masses.

Finally, we consider the distribution in the cosine of the angle
between the positron and the anti-muon in
\reffi{plot:cosine_angle_separation_epmu}.  The absolute prediction
nicely illustrates that the charged leptons produced via the
QCD-induced mechanism are mainly back to back while the EW-induced
process has a maximum both for the back-to-back and the collinear
configurations (the drop in the last bin is due to the cut on $\Delta R_{\ell\ell}$).  
The latter arise from configurations with a
strongly boosted VBS system that does not occur in QCD-induced
topologies. The observable is thus an example of a relatively
inclusive quantity where the ratio of the VBS and QCD-induced contributions vary in shape in phase space.
While the  QCD-induced contributions  are strongly suppressed for
small and intermediate angles, they are of the same order of magnitude
as the EW contributions for very large angles.
The interference contribution is relatively
constant over the whole spectrum and strongly suppressed.  
{The $\mathcal{O}{\left(\alpha^{7}\right)}$ and
$\mathcal{O}{\left(\alphas\alpha^{6}\right)}$ corrections both
vary steadily with increasing $\cos\theta_{\Pe^+\mu^-}$ 
from $-11\%$ to $-14\%$ and from $-3\%$ to $-4\%$, respectively.
The corrections are of the same order as for the fiducial cross section.
The $\mathcal{O}{\left(\alphas^{3}\alpha^{4}\right)}$ contribution is
small for small angles but increases in size to 
$-6\%$ 
for large
angles, where it lies between the
$\mathcal{O}{\left(\alphas\alpha^{6}\right)}$ and the
$\mathcal{O}{\left(\alpha^{7}\right)}$ corrections.}
This is due to the
enhanced LO QCD-induced contribution and
confirms that
the hierarchy of the NLO corrections is determined to a large extent
by the pattern observed for the LO prediction.
{The contribution
at order $\mathcal{O}{\left(\alphas^2\alpha^{5}\right)}$ has
qualitatively a similar behaviour, it is however suppressed.}

\begin{figure}
        \setlength{\parskip}{-10pt}
        \begin{subfigure}{0.49\textwidth}
                \subcaption{}
                \includegraphics[width=\textwidth]{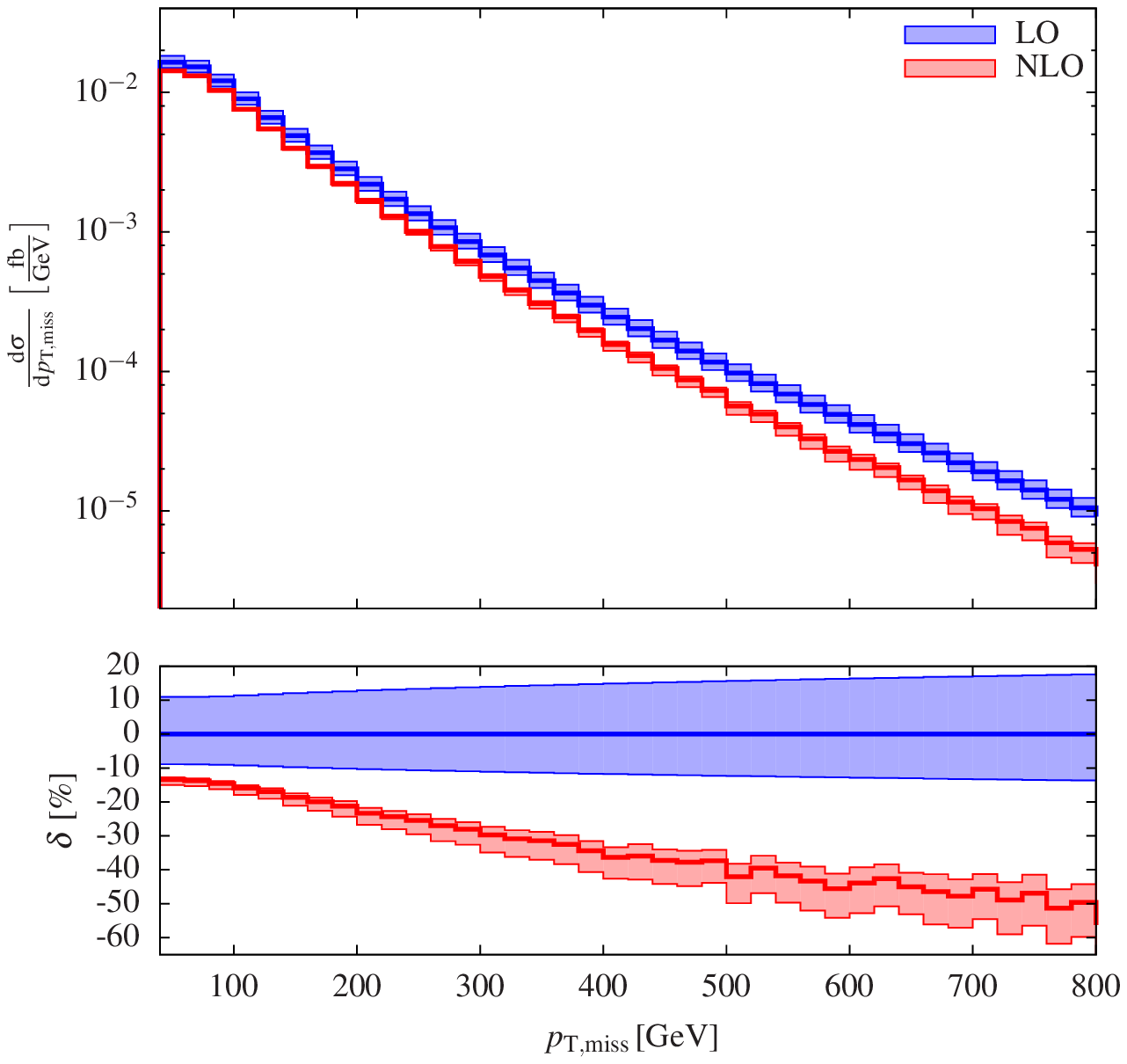}
                \label{plot:transverse_momentum_truth_missing_scale}
        \end{subfigure}
        \hfill
        \begin{subfigure}{0.49\textwidth}
                \subcaption{}
                \includegraphics[width=\textwidth]{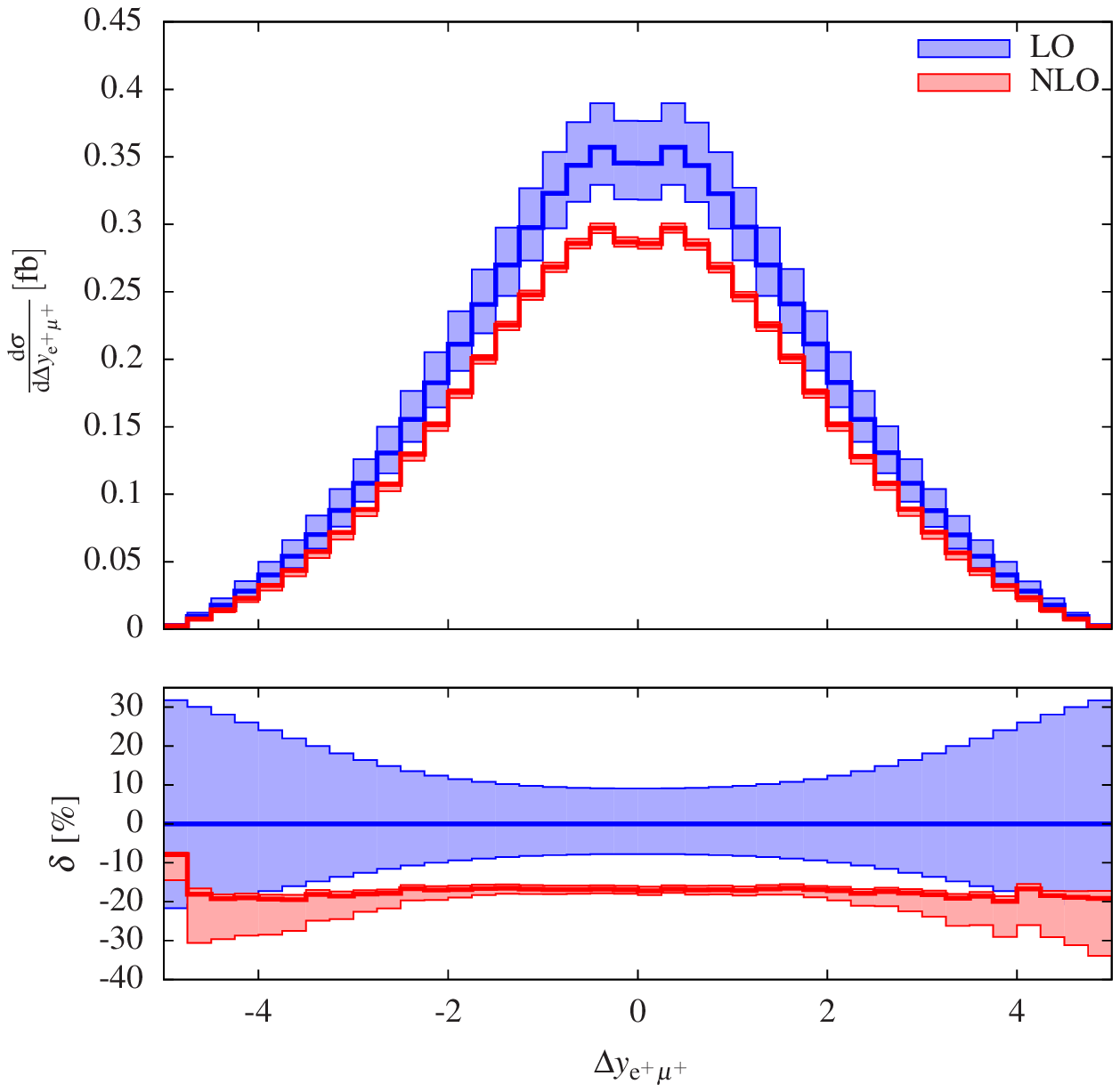}
                \label{plot:rapidity_separation_pomu_scale} 
        \end{subfigure}

        \begin{subfigure}{0.49\textwidth}
                \subcaption{}
                \includegraphics[width=\textwidth]{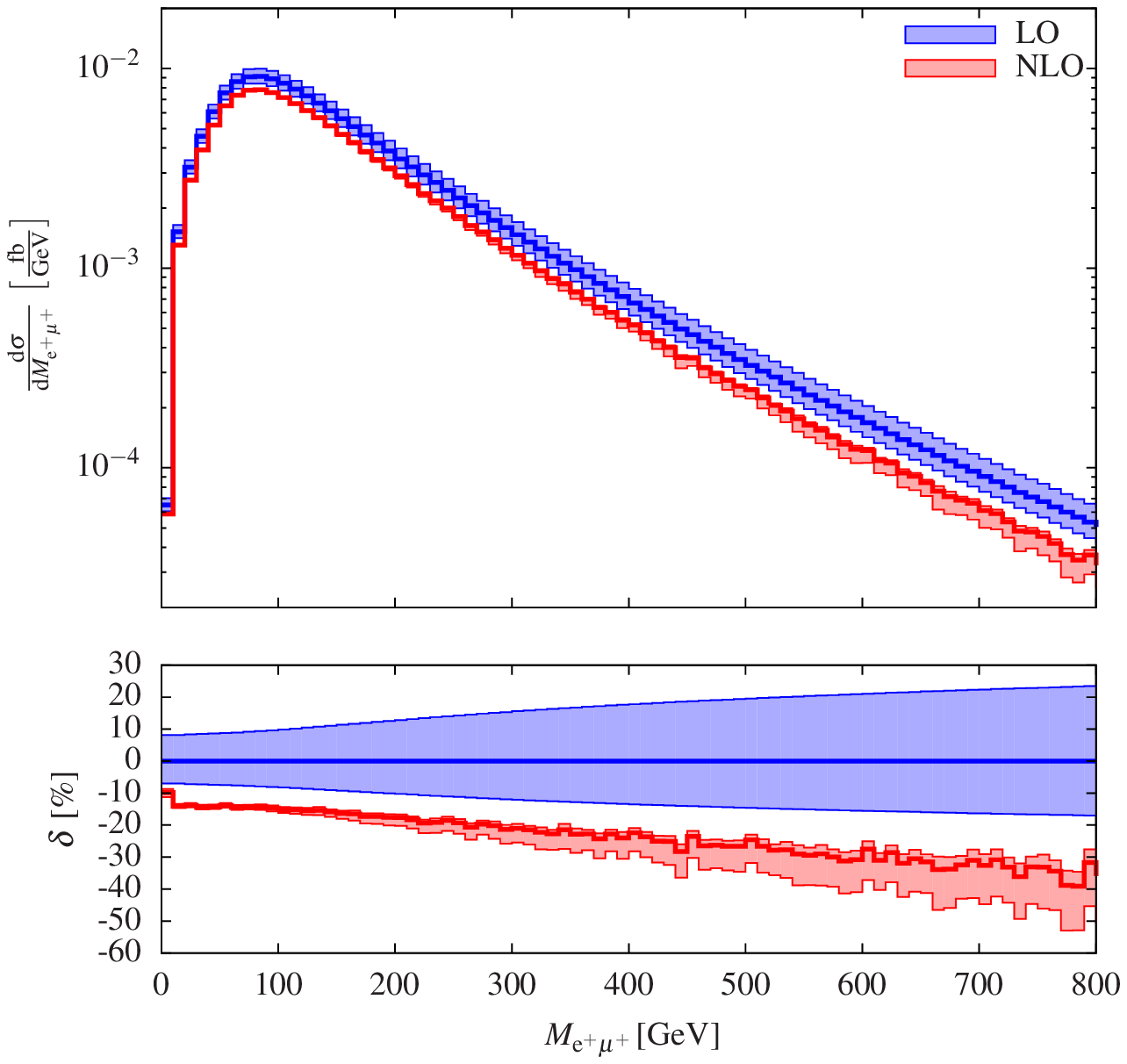}
                \label{plot:invariant_mass_epmu_scale}
        \end{subfigure}
        \hfill
        \begin{subfigure}{0.49\textwidth}
                \subcaption{}
                \includegraphics[width=\textwidth]{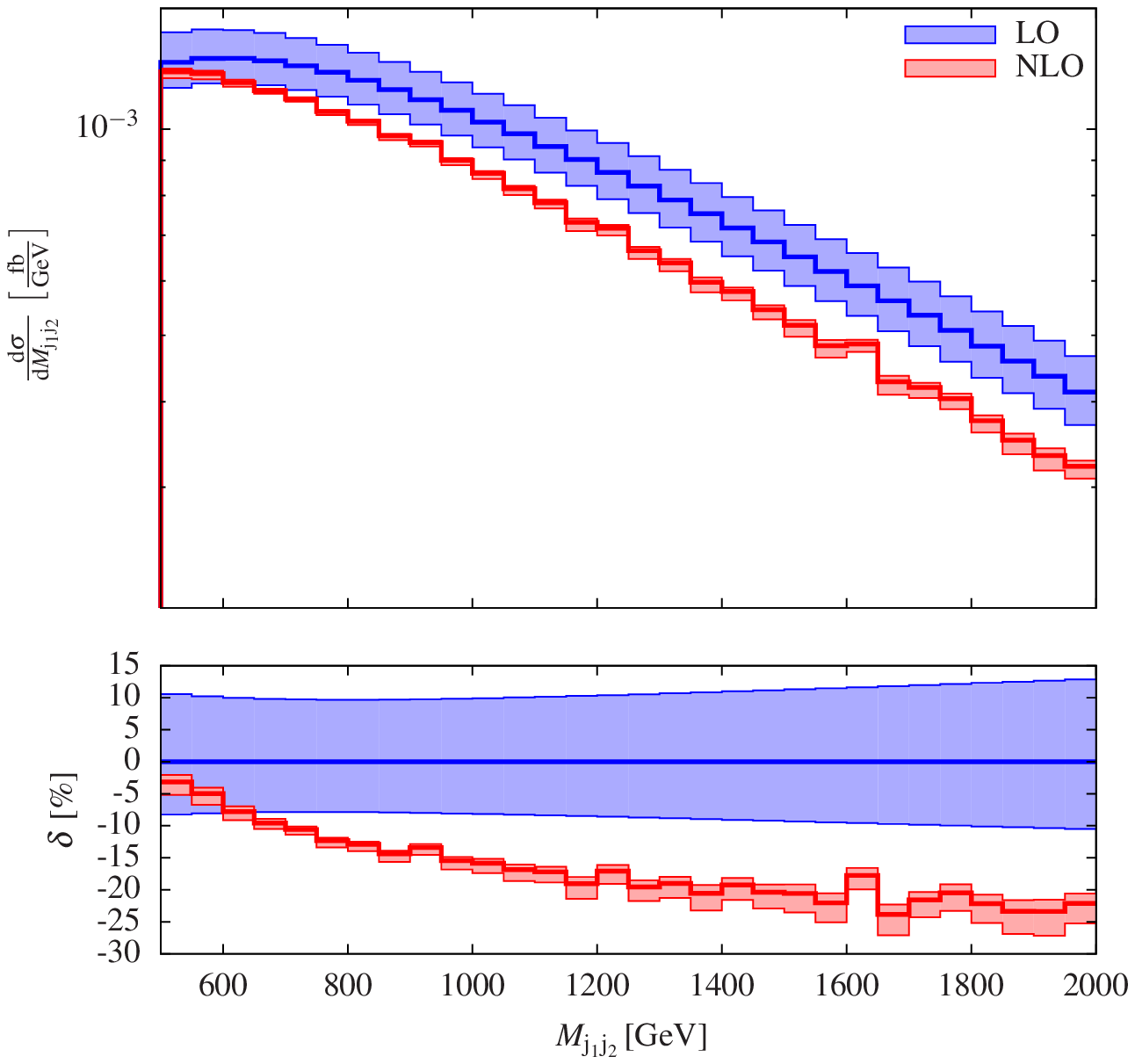}
                \label{plot:invariant_mass_mjj12_scale} 
        \end{subfigure}
        
        \vspace*{-3ex}
        \caption{\label{fig:distributions_scale}%
                Differential distributions at a centre-of-mass energy $\sqrt{s}=13\TeV$ at the LHC for $\Pp\Pp\to\mu^+\nu_\mu\Pe^+\nu_{\Pe}\Pj\Pj$: 
                \subref{plot:transverse_momentum_truth_missing_scale}~missing transverse momentum~(top left), 
                \subref{plot:rapidity_separation_pomu_scale}~rapidity separation between the positron and anti-muon~(top right),
                \subref{plot:invariant_mass_epmu_scale}~invariant mass
                of the positron and anti-muon system~(bottom left), 
                \subref{plot:invariant_mass_mjj12_scale}~invariant mass of the two tagging jets~(bottom right).
                The upper panels show the sum of all LO and NLO contributions with scale variation.
                The lower panels show the relative corrections
                in per cent.  }
\end{figure}

In \reffi{fig:distributions_scale} we present some
distributions displaying the variation of the factorisation and
renormalisation scales.  
In the upper panels, the sums of all LO contributions as well as of all
NLO contributions are shown.\footnote{The photon-induced contributions
  are left out of the NLO predictions.}  The band is obtained by
varying the 
factorisation and renormalisation scales independently by the factors
$\xi_{\rm fac}$ and $\xi_{\rm ren}$ with the combinations of
Eq.~\eqref{eq:scales}. The central scale is defined as $(\xi_{\rm fac},\xi_{\rm ren}) =
\left(1,1\right)$.  The relative
corrections shown in the lower panel are normalised to the LO prediction for the
central scale.  In
\reffi{plot:transverse_momentum_truth_missing_scale}, the distribution
in the missing transverse momentum is displayed.  The
scale-uncertainty band  decreases significantly by going from LO to NLO.
Nonetheless, these two bands do not overlap.  Indeed, as
explained previously, the VBS process (which is a purely EW process) is
dominating the $\mu^+\nu_\mu\Pe^+\nu_{\Pe}\Pj\Pj$ final state, and the NLO EW corrections to VBS
represent a large fraction of the NLO corrections.  These corrections have the
effect of simply shifting the prediction without affecting
significantly the size of the scale variation band.  
While missing higher-order QCD corrections can be estimated via
scale variations this is not possible for higher-order EW corrections
in the on-shell scheme. A conservative estimate for the
higher-order EW corrections is provided by the square of the EW NLO
correction, $(\delta_{\rm EW})^2$.

In \reffi{plot:rapidity_separation_pomu_scale}, the distribution in
the rapidity difference of the positron and the anti-muon is shown.
As for the rapidity of the anti-muon, the bulk of the cross section is
located in the central region due to the dominance of the VBS process
in this region.  For large rapidities, where the QCD-induced background
contributions are sizeable, the LO scale variation is
particularly large and the LO and NLO uncertainty bands overlap.

Finally, the distributions in the invariant masses of the
positron--anti-muon system and of the two tagging jets are shown in
\reffis{plot:invariant_mass_epmu_scale} and
\ref{plot:invariant_mass_mjj12_scale}, respectively.
These two observables have been considered 
in {a} recent CMS measurement
\cite{CMS:2017adb}. The behaviour of the NLO corrections and the scale
dependence is similar as for the distribution in the missing
transverse momentum in
\reffi{plot:transverse_momentum_truth_missing_scale}.
Note, however, that the NLO corrections are larger for the
distributions in $p_{\rm T,miss}$ and $M_{\Pe^+\mu^+}$ than for the
one in  $M_{\Pj\Pj}$.
In general, the scale dependence is larger where the cross section is
smaller and the NLO corrections are larger.

\section{Conclusions}
\label{sec:Conclusions}

In this article we have presented all NLO electroweak (EW) and QCD
corrections to the process $\Pp\Pp\to\mu^+\nu_\mu\Pe^+\nu_{\Pe}\Pj\Pj$
including like-sign charged vector-boson scattering (VBS) and its
EW- and QCD-induced irreducible background.  As the
full LO and NLO matrix elements are used, these computations account
for all possible off-shell, non-resonant, and interference effects.
The latter aspect plays an important role in this computation: the LO
amplitude consists of a purely 
EW-induced part, which includes
VBS, and a QCD-induced part, leading thus to three different LO
contributions at the level of squared amplitudes. These are of the
orders $\mathcal{O}{\left(\alpha^{6}\right)}$,
$\mathcal{O}{\left(\alphas\alpha^{5}\right)}$,
$\mathcal{O}{\left(\alphas^{2}\alpha^{4}\right)}$ in the strong
and electromagnetic couplings.  At NLO, consequently, four types of
corrections have been computed at the orders
$\mathcal{O}{\left(\alpha^{7}\right)}$,
$\mathcal{O}{\left(\alphas\alpha^{6}\right)}$,
$\mathcal{O}{\left(\alphas^{2}\alpha^{5}\right)}$, and
$\mathcal{O}{\left(\alphas^{3}\alpha^{4}\right)}$, respectively.
For the orders
$\mathcal{O}{\left(\alphas\alpha^{6}\right)}$ and
$\mathcal{O}{\left(\alphas^{2}\alpha^{5}\right)}$, both NLO QCD and EW
corrections to different underlying Born contributions arise.  These
cannot be unambiguously separated 
as some loop diagrams contribute to
both.
Hence, at NLO, it is not possible to strictly distinguish
the VBS process from its irreducible QCD background processes.

We have presented predictions for the LHC running at a
centre-of-mass energy of $13\TeV$ with realistic experimental event
selections applied to the final state.  In this fiducial region,
results for the integrated cross-section and various distributions have been shown.  In particular, predictions for the
three LO contributions as well as for the four contributing NLO
corrections have been presented both separately and in a combined
form.  This allows the experimental physicists to extract all
necessary information from our calculation and to include it in their
analysis.

At LO, the VBS process clearly dominates over its irreducible
background processes.  On the one hand, this is due to the
characteristic signature of two equally charged W bosons excluding a
sizeable amount of partonic channels that would mainly contribute to
the QCD background. On the other hand, it is further enhanced by the
specific VBS event selection.  Concerning the NLO corrections, we
identify the dominant contributions to be the large negative EW
corrections to the VBS process.
For the fiducial cross section, they reach $-13\%$ of the complete LO contributions 
and are even significantly more enhanced at the level of differential distributions with up to (minus) $40\%$ corrections in the kinematical regions explored.
These corrections display the typical behaviour of Sudakov logarithms that
grow large in the high-energy regime.  
The NLO contributions of order
$\mathcal{O}{\left(\alphas\alpha^{6}\right)}$, which are dominated by
QCD corrections to the EW-induced process, are four times smaller and
negative for the fiducial cross section.  At the level of differential
distributions, they display a different behaviour than the EW
corrections.  Finally, the 
NLO contributions
of order $\mathcal{O}{\left(\alphas^{2}\alpha^{5}\right)}$ and
$\mathcal{O}{\left(\alphas^{3}\alpha^{4}\right)}$ are relatively
suppressed with respect to the LO prediction and even cancel
partially.  The dependence on the factorisation and renormalisation
scale is significantly reduced upon including NLO corrections.
However, this does not provide an estimate of the theoretical
uncertainty from missing higher-order EW corrections.  Since these are
dominated by EW Sudakov logarithms, we recommend to use the
squared EW corrections as a conservative estimate for this purpose.

As this article presents predictions for a realistic final state
where the event selection follows the one of the experimental
collaborations, this should make this computation very relevant for
the measurement of the VBS process.  Since at NLO
it is not possible to distinguish unambiguously the VBS process from
its irreducible background, we advocate for a global measurement
of the $\mu^+\nu_\mu\Pe^+\nu_{\Pe}\Pj\Pj$ final state.

\acknowledgments We thank Jean-Nicolas Lang and Sandro Uccirati for
supporting the computer program \recola\ and Robert Feger
for assistance with the Monte Carlo program \mocanlo.
We acknowledge financial support by the German Federal Ministry for Education and
Research (BMBF) under contract no.~05H15WWCA1 and the German Science
Foundation (DFG) under reference number DE 623/6-1.

\bibliographystyle{JHEPmod}
\bibliography{vbs_nlo} 

\end{document}